\documentclass[english, 12pt]{article}
\usepackage[bookmarks=true, colorlinks=true, linkcolor=blue, citecolor=blue, breaklinks=true]{hyperref}
\usepackage{natbib}
\usepackage[T1]{fontenc}
\usepackage[active]{srcltx}
\usepackage{booktabs}
\usepackage{babel}
\usepackage{float}
\usepackage{amsthm}
\usepackage{amsmath}
\usepackage{amssymb}
\usepackage{graphicx, graphics}
\usepackage{setspace}
\usepackage{esint}
\usepackage{algorithm}
\usepackage{subfig}
\usepackage{changepage}

\usepackage[usenames]{color}
\usepackage[normalem]{ulem}
\definecolor{mygreen}{rgb}{0,.5,0}

\newcommand{\M}{{\mathcal{M}}}
\newcommand{\cS}{{\mathcal{S}}}

\newcommand{\R}{{\mathbb{R}}}

\newcommand{\be}{\begin{equation}}
\newcommand{\ee}{\end{equation}}
\newcommand{\bee}{\begin{equation*}}
\newcommand{\eee}{\end{equation*}}

\newtheorem{theorem}{Theorem}[section]

\graphicspath{{../figures/}}

\theoremstyle{plain}
  
  \theoremstyle{plain}
  
  \theoremstyle{plain}
  \newtheorem*{assumption*}{\protect\assumptionname}
  \theoremstyle{plain}
  
  \theoremstyle{remark}
  \newtheorem*{rem*}{\protect\remarkname}
  \theoremstyle{plain}

\usepackage{fullpage}
 \usepackage{pdfsync}
\usepackage[title,titletoc]{appendix}

  \providecommand{\assumptionname}{Assumption}
  \providecommand{\lemmaname}{Lemma}
  \providecommand{\propositionname}{Proposition}
  \providecommand{\remarkname}{Remark}
\providecommand{\theoremname}{Theorem}
\providecommand{\corollaryname}{Corollary}

\begin{document}
\title{A Machine Learning Algorithm for Finite-Horizon
Stochastic Control Problems in Economics\thanks{Xianhua Peng is partially supported by the Natural Science Foundation of Shenzhen (Grant No. JCYJ20190813104607549) and the National Natural Science Foundation of China (Grant No. 72150003). The paper was previously entitled ``EM Algorithm and Stochastic Control."}}

\date{This version: December 6, 2024}

\author{Xianhua Peng\thanks{HSBC Business School, Peking University, University Town, Nanshan District, Shenzhen, 518055, China. Email: xianhuapeng@pku.edu.cn.}  
\and Steven Kou\thanks{Questrom School of Business, Boston University, Rafik B. Hariri Building, 595 Commonwealth Avenue, Boston, MA 02215, USA. Email: kou@bu.edu.} 
\and  Lekang Zhang\thanks{HSBC Business School, Peking University, University Town, Nanshan District, Shenzhen, 518055, China. Email: lekang\_zhang@stu.pku.edu.cn.}
		}

\maketitle

\begin{abstract}  
\begin{adjustwidth}{-0.2cm}{-0.2cm}
We propose a machine learning algorithm for solving finite-horizon stochastic control problems based on a deep neural network representation of the optimal policy functions.  The algorithm has three features: (1) It can solve high-dimensional (e.g., over 100 dimensions) and finite-horizon time-inhomogeneous stochastic control problems. (2) It has a monotonicity of performance improvement in each iteration, leading to good convergence properties. (3) It does not rely on the Bellman equation. To demonstrate the efficiency of the algorithm, it is applied to solve various finite-horizon time-inhomogeneous problems including recursive utility optimization under a stochastic volatility model, a multi-sector stochastic growth, and optimal control under a dynamic stochastic integration of climate and economy model with eight-dimensional state vectors and 600 time periods.


    \vspace{0.5em}

\end{adjustwidth}

\end{abstract}

\baselineskip 18pt

\section{Introduction}

Stochastic control problems are widely used in macroeconomics (e.g., the study of stochastic growth and real business cycle), microeconomics (e.g., utility maximization problem), and finance (e.g., portfolio choices and optimal execution).
Indeed, there is a large literature on stochastic control in economics. For example, \citet*{stokey1989recursive}  describe many economic models using stochastic control, including economic growth, resource extraction, principal-agent problems, business investment, asset pricing, etc. 
\citet*{Hansen-Sargent-2013} give detailed discussions on stochastic control problems in which
the Bellman equations can be solved analytically, especially problems with quadratic objective functions and linear transition functions. \citet*{Ljungqvist-Sargent-2018_updated} discuss dynamic programming methods and their applications to a variety of problems in economics.
\citet*{miao2020economic} gives a comprehensive introduction to the analytical and numerical tools for solving stochastic control problems in economics.

Despite the previous efforts, three significant obstacles remain: (i) Many stochastic control problems in economics are finite-horizon time-inhomogeneous problems, which may be more difficult than the related infinite horizon problem, as the optimal control policies at different time periods are different. 
(ii) Due to the curse of dimensionality, it is generally difficult to numerically solve stochastic control problems in high dimensions and for problems with complicated stochastic dynamics. 
(iii) If the utility function in the control problem is not time-separable, then such a problem may not have the Bellman equation.

To overcome these difficulties, we propose a machine learning algorithm, the Monotonic Monte Carlo Control (MMCC) algorithm,
to solve high-dimensional, finite time horizon, and time-inhomogeneous stochastic control problems without using dynamic programming principles. The MMCC algorithm can be implemented using a deep neural network representation of the policy functions where the parameters of the neural networks can be learned by stochastic gradient descent approach.
In each round of training, the algorithm first generates sample paths of the states and controls by Monte Carlo simulation, and then 
updates the control policies in each time period sequentially in a backward direction. 
Therefore, The MMCC algorithm has a monotonicity of performance improvement in each iteration step, leading to good convergence properties.
The algorithm does not require the Bellman equation, does not require the utility function to be time-separable, and allows general stochastic dynamics of the evolution of states.
We demonstrate the effectiveness of the MMCC algorithm by solving various high-dimensional (e.g., over 100-dimensional) stochastic control problems including portfolio selection under a stochastic volatility model, multi-sector stochastic growth, and optimal control under a dynamic stochastic integration of climate and economy model with 8-dimensional state vectors and 600 time periods. 

\subsection{Literature Review}

\citet*{Judd-1998} and \citet*{Miranda-Fackler-2002} provide a comprehensive treatment of traditional numerical methods such as value function iteration for stochastic control problems in economics. 
Recently, some grid point-based and grid point-free machine learning methods have been proposed for solving discrete-time dynamic economic models. 

First, for infinite-horizon and representative-agent (RA) problems, \citet*{Lepetyuk2020} employ supervised neural networks to learn the policy function at grid points defined through unsupervised clustering analysis, thereby using a grid-based method.
Within the functional iteration framework, supervised learning is typically applied. For example, \citet*{Renner-Scheidegger-2018} and \citet*{scheidegger-Bilionis-2019} use Gaussian process regression to approximate the value function; \citet*{valaitis2024machine} use neural networks to approximate the expectation term in the Euler equation. Outside the framework of functional iteration, \citet*{maliar2021deep} provide a unified unsupervised framework encompassing three optimization strategies: maximizing lifetime reward, minimizing equilibrium-condition error (i.e., Euler-equation error), and minimizing Bellman-equation error.  In the case of Euler-equation error minimization, \citet*{Pascal2024_published} generalizes the Monte Carlo operator in \citet*{maliar2021deep}. Additionally, in the case of continuous-time infinite-horizon RA problems, \citet*{duarte2024machine} use two neural networks to approximate the policy and value function and adopt supervised learning to train the two neural networks to solve the Hamilton-Jacobi-Bellman (HJB) equation.

For infinite-horizon and heterogeneous-agent (HA) problems, grid-free methods based on simulation are proposed. \citet*{maliar2021deep} demonstrate the effectiveness of their grid-free framework by solving two representative-agent models and one heterogeneous-agent model \citep*{krusell-smith-1998}.
Additionally, \citet*{Han2021} employ both supervised and unsupervised learning using two neural networks in a grid-free manner, with unsupervised learning applied to policy function approximation and supervised learning to value function approximation. They essentially aim to maximize lifetime reward and then evaluate their method based on Bellman-equation error.
Moreover, \citet*{HallHoffarth2023} minimizes the error of equilibrium conditions to solve a heterogeneous-agent New Keynesian (HANK) model. In the case of continuous-time infinite-horizon HA problems, \citet*{huang2023probabilistic} focuses on the probabilistic formulation of economic models and uses deep learning to solve the system of forward-backward stochastic differential equations corresponding to the model. Subsequently, \citet*{huang2023breaking} employs the probabilistic approach to solve a continuous-time version of \citet*{krusell-smith-1998} and an asset pricing model with search-and-bargaining frictions.

For finite-horizon problems, existing literature focuses on solving overlapping generations models by grid-free methods. For example, \citet*{duarte2021simple} solve a life-cycle model by maximizing lifetime reward; \citet*{azinovic2022deep} and \citet*{Azinovic2023} solve an overlapping generations model by minimizing the error of equilibrium conditions. In particular, \citet*{Azinovic2023} introduce a market clear layer to enforce the economic constraints.

This paper complements the existing literature in four aspects: (i) The proposed MMCC algorithm is a grid point-free (i.e.,  simulation-based) algorithm, which avoids the curse of dimensionality and enables us to handle high dimensional problems (e.g., 100 dimensions). 
In this regard, our algorithm is closely related to the problem of American option pricing using simulation (see, e.g., \cite*{Longstaff2001}, \citet*{Tsitsiklis-VanRoy-2001}, and \citet*{Broadie-Glasserman-1997,  Broadie-Glasserman-2004}, \citet[][Ch. 8]{Glasserman-2004}). (ii) The MMCC algorithm differs from existing grid point-free methods in how network parameters update. In each round of iteration, the MMCC algorithm updates the neural networks of the control policies at different time periods sequentially in a backward manner. In contrast, all neural network parameters are updated simultaneously in existing grid point-free methods.
(iii) The MMCC algorithm has a monotonicity of performance improvement at each iteration, while many existing algorithms do not have such a property.   (iv) The MMCC algorithm does not use the Euler equation or Bellman equation; in contrast, many numerical algorithms and grid point-free algorithms in the literature rely on the Euler equation, Bellman equation, or their approximation. 
(v) The MMCC algorithm can solve problems with time-inseparable utility functions, which may not have Bellman equations.

Besides the economic literature, there is a large body of literature on applied mathematics and applied probability in stochastic control.  In particular, 
\citet*{Yong-Zhou-1999} and \citet*{Flemming-Soner-2006} provide an in-depth discussion on continuous time stochastic control problems and their applications. \citet*{Kushner-Dupuis-2001} give an excellent survey of numerical methods for solving continuous time stochastic control problems by using Markov chains. There have also been many studies on the numerical solutions to continuous time stochastic control problems in mathematical finance.\footnote{See, e.g., \cite*{ZHANG2004}, \cite*{Bouchard2004}, 
	\cite*{chass2014}, \cite*{chass2016},  \cite*{Crisan2010}, \citet*{gobet2005}, \citet*{gobet2016, gobet2017}, \citet*{henry2014},  \citet*{henry2019_published},  
	\cite*{Kharroubi2015_published}, \cite*{Kharroubi2014_published}, and \citet*{GZZ-2015_published}.} Most of these studies focus on particular stochastic processes, e.g., discretized diffusion processes or L\'{e}vy processes, but our MMCC algorithm can be applied to general stochastic processes. Moreover, our method is a simulation-based method, suitable for high-dimensional problems.

Approximate dynamic programming (ADP) has been developed\footnote{ADP has also evolved under the name of reinforcement learning in computer science (see, e.g.,  \citet*{Sutton-Barto-1998}).} for dealing with three sources of curses of dimensionality: high dimensionality of state space, control policy space, and random shock space; see the books by \citet*{Powell-2011} and \citet*{Bertsekas-2012}.
ADP algorithms can be broadly classified into two categories: value iteration and policy iteration.\footnote{Many ADP algorithms focus on infinite time horizon problems where the optimal value
function and policy are stationary. In contrast, our MMCC algorithm focuses on finite time horizon problems where neither the optimal value function nor the optimal policy is stationary.} 
Most ADP algorithms are value iteration algorithms,
which approximate the value function by employing the Bellman equation.\footnote{Value function iteration is closely related to the duality approach for stochastic dynamic programming; see, e.g., \citet*{Brown-Smith-Sun-2010}, \citet*{Brown-Smith-2014}, \citet*{Brown-Haugh-2017_published}, and \cite*{Chen-Ma-Yu-2016}.}
As an alternative, a policy iteration algorithm keeps track of
the policy instead of the value function. At each period, a value function is calculated based on a policy estimated previously and then improved within the policy space. 
However, the value iteration and policy iteration ADP algorithms may not have a monotonic improvement of the value function at each iteration.
The MMCC algorithm is related to but is fundamentally different from the policy iteration ADP algorithms mainly in that: (i) The MMCC algorithm does not use the Bellman equation; (ii) The MMCC algorithm has a monotonic improvement of the value function at each iteration; (iii) The MMCC algorithm can be applied to general control problems in which the objective functions may not be time-separable.

Deep neural networks were first used in \citet*{HanE2016} to solve stochastic control problems. They solve finite-horizon problems by approximating the time-dependent controls as feedforward neural networks at each time period; see further extensions on solving partial differential equations and stochastic differential equations in \citet*{EHan2017_published} and \citet*{BeckEJ2019_published}. Additionally, \citet*{Reppen2023} leverage this algorithm to solve high-dimensional problems in American and Bermudan option pricing. \citet*{Hure-Pham-Bachouch-Lang-2021} propose solving finite-horizon stochastic control problems based on dynamic programming (i.e., the Bellman equation). They use two neural networks at each time period: one for representing the control policy and the other for representing the value function; see further numerical applications of this algorithm in \citet*{Bachouch_2021}.
The MMCC algorithm differs from these papers in two aspects: (i) The MMCC algorithm leads to monotonic improvement of the value function in each iteration, while these algorithms do not.
(ii) We provide applications of the MMCC algorithm to solve various economic problems such as multi-sector stochastic growth and the social cost of carbon emission problem. 

The literature on Markov decision processes mainly concerns multi-period stochastic control problems with a finite state space or a finite control space. There are also simulation-based algorithms for Markov decision processes; see, e.g., the books by \citet*{chang2013simulation} and \citet*{Gosavi-2015} for comprehensive review and discussion. The main differences between these algorithms and our MMCC algorithm are: (i) The MMCC algorithm has monotonicity in each iteration; (ii) The MMCC algorithm does not use the Bellman equation.

The rest of the paper is organized as follows. The algorithm is proposed in Section \ref{sec:cem_algo}, and in Section \ref{sec:Convergence-Analysis} we show that the algorithm improves the objective function monotonically in each iteration and hence has good convergence properties. In Section \ref{sec:Simulation_Based_Algorithm}, we propose an implementation of the algorithm via deep neural network approximation of the policy functions.  
To update the policy functions, one can use stochastic gradient descent 
in each iteration. The applications of the MMCC algorithm to solve the recursive utility optimization problem under a stochastic volatility model, 
multi-sector stochastic growth problem, and the problem of the social cost of carbon are given in Sections \ref{sec:recursiveUtility}, \ref{sec:business_cycle}, and \ref{sec:scc} respectively.

\section{The MMCC Algorithm}\label{sec:cem_algo}
\subsection{The Setting of the Problem}

We consider a general finite time horizon stochastic control problem,  with $T$ periods from $0$ to $T-1$.
Let $n_c$ be the dimension of the control policy and let $n_s$ be the dimension of the state. 
At the $t$-th period the decision maker observes the state $s_t\in \mathbb{R}^{n_s}$ and then chooses a $n_c$-dimensional control $c_t\in\sigma(s_t)$, the sigma field generated by $s_t$. Hence,
the policy $c_{t}$ is adapted to the information available up to period $t$ and can be represented as a function of $s_t$. The initial state $s_0$ is given at period $0$.
The state $s_{t+1}$ is determined by $s_t$, $c_t$, and random shock by the following state evolution equation
\begin{equation}\label{eq:state_evo}
s_{t+1}=\psi_{t+1}(s_{t},c_{t},z_{t+1}),~~0\leq t\leq T-1,~~~~s_{0} \text{ is given},
\end{equation}
where $\psi_{t+1}(\cdot)$ is the state evolution function and
$z_{t+1}\in\mathbb{R}^{n_z}$ is the random vector denoting the random shock in the $(t+1)$th period.
Path dependence (i.e., $c_t$ may depend on states $s_k$ for some $k<t$) can be accommodated by including auxiliary
variables in $s_t$.
The state evolution dynamics in \eqref{eq:state_evo} is a general one, which is not restricted to discretized diffusion processes or L\'{e}vy
processes.

The goal is to find the optimal control policy.
For $t\geq 1$, we assume that the control policy can be represented as
\begin{equation}\label{eq:c_t_s_t}
c_t = c(t, s_t, \theta_t), t\geq 1,
\end{equation}
where $c(\cdot)$ is a function and $\theta_t=(\theta_{t, 1}, \theta_{t, 2}, \ldots, \theta_{t, d})^\top \in \mathbb{R}^d$ is the vector of parameters for the $t$th period. 

The policy function 
$c(t,\cdot, \theta_t)$ is to be determined and can be represented by a deep neural network with parameter $\theta_t$. 
More precisely, consider a deep neural network with $k$ hidden layers, then the function $c(t,\cdot, \theta_t)$ can be written a composite function
$$\sigma_{k} \circ T_{k} \circ \cdots \sigma_{2} \circ T_2 \circ \sigma_1 \circ T_1,$$ 
where $T_i$ is an affine function and $\sigma_i$ is a nonlinear activation function, such as the rectified linear unit $\sigma (x)=\max(x,0)$.

In the case of deep neural networks, the control problem amounts to finding the affine functions $T_i$, $i \geq 1$, and $c_0$. This is possible by using stochastic gradient descent algorithms such as Adam \citep{KingmaBa} if the stochastic gradient of the objective function can be found analytically.

At period 0, the decision maker wishes to choose the optimal control $c_0\in\mathbb{R}^{n_c}$ and the sequence of control parameters $\theta_1, \ldots, \theta_{T-1}$, which determines the sequence of controls $c_1, \ldots, c_{T-1}$, to maximize the expectation of his or her utility
{\allowdisplaybreaks
	\begin{align}
	\max_{(c_0, \theta_{1}, \ldots, \theta_{T-1})\in \Theta}\ \ & \mathbb{E}_0\left[\sum_{t=0}^{T-1}u_{t+1}(s_{t+1}, s_t, c_{t})\middle | c_0, \theta_{1}, \ldots, \theta_{T-1}\right]\label{equ:multi_per_obj}\\
	\text{s.t.}\,\ \ \ \ \ \ \ \  & c_t = c(t, s_t, \theta_t), t = 1, \ldots, T-1,\label{equ:control}\\
	& s_{t+1}=\psi_{t+1}(s_{t},c_{t},z_{t+1}), ~~0\leq t\leq T-1,~~s_{0}\text{ is given}, \notag
	\end{align}}
where $\Theta$ is a subset of $\mathbb{R}^{n}$ with $n=n_c+(T-1)d$; $u_{t+1}(\cdot)$ is the utility function of the decision maker in the $(t+1)$th period. It is worth noting that the utility function in the first period can include utility at period $0$. 

A control problem more general than the problem \eqref{equ:multi_per_obj} is given by
{\allowdisplaybreaks
	\begin{align}
	\max_{(c_0, \theta_{1}, \ldots, \theta_{T-1})\in \Theta}\ \ & \mathbb{E}_0\left[u(s_0, c_0, s_1, c_1, \ldots, s_{T-1}, c_{T-1}, s_T)\middle | c_0, \theta_{1}, \ldots, \theta_{T-1}\right]\label{equ:multi_per_obj_gen}\\
	\text{s.t.}\,\ \ \ \ \ \ \ \  & c_t = c(t, s_t, \theta_t), t = 1, \ldots, T-1,\label{equ:state_evol_gen}\\
	& s_{t+1}=\psi_{t+1}(s_{t},c_{t},z_{t+1}), ~~0\leq t\leq T-1,~~~s_{0}\text{ is given}, \notag
	\end{align}}
where $u(s_0, c_0, s_1, c_1, \ldots, s_{T-1}, c_{T-1}, s_T)$ is a general utility function that may not be time-separable as the one in \eqref{equ:multi_per_obj}. For simplicity of exposition, we will present our MMCC algorithm for the problem \eqref{equ:multi_per_obj}; however, the MMCC algorithm also applies to the general problem \eqref{equ:multi_per_obj_gen}; see Appendix \ref{app:CEM-general-control} for details.

For simplicity of notation, we denote $x=(c_0, \theta_1, \theta_2, \ldots, \theta_{T-1})$ and denote
the objective function of problem \eqref{equ:multi_per_obj} by
\be\label{equ:utility_func}
U(x):=U(c_0, \theta_1, \theta_2, \ldots, \theta_{T-1}):=\mathbb{E}_0\left[\sum_{t=0}^{T-1}u_{t+1}(s_{t+1}, s_t, c_t)\middle | c_0, \theta_{1}, \ldots, \theta_{T-1}\right].
\ee
In general, the expectation in \eqref{equ:utility_func} cannot be evaluated in closed form, and hence $U(x)$ does not have an analytical form.

\subsection{Description of the MMCC Algorithm}

The MMCC algorithm is an iterative algorithm for solving \eqref{equ:multi_per_obj}, involving multiple rounds of the back-to-front updates, that updates the control policy at a given time period by optimizing the objective function with respect to the control policy at that time period only, and with
the control policies at all other periods fixed at their most up-to-date status in the iteration of the algorithm.

More precisely, suppose that after the $(k-1)$th iteration, the control policy parameter is $x^{k-1}:=(c^{k-1}_0, \theta^{k-1}_{1}, \theta^{k-1}_{2}, \ldots, \theta^{k-1}_{T-1})$. In the $k$th iteration, the MMCC algorithm updates $x^{k-1}$ to be $x^{k}:=(c^{k}_0, \theta^{k}_{1}, \theta^{k}_{2}, \ldots, \theta^{k}_{T-1})$ by the updating rule:
\be\label{equ:def_pts_map}
x^k\in M(x^{k-1}),
\ee
where $M(\cdot)$ is a point-to-set map on $\Theta$ (i.e., $M(\cdot)$ maps a point in $\Theta$ to a subset of $\Theta$) that represents the updating rule. 
At each time period $t=T-1, T-2, \ldots, 1$, the algorithm updates $\theta^{k-1}_{t}$ to be $\theta^{k}_{t}$ and then moves backward to update $\theta^{k-1}_{t-1}$; at last, the algorithm updates $c^{k-1}_0$ to be $c^{k}_0$.

Next, we specify the precise updating rule in \eqref{equ:def_pts_map}. In the $k$th iteration, before updating the control parameter at period $t\in \{T-1, T-2, \ldots, 1\}$, the control policy parameter is $(c^{k-1}_0, \theta^{k-1}_{1}, \ldots, \theta^{k-1}_{t-1}, \theta^{k-1}_{t}, \theta^{k}_{t+1}, \theta^{k}_{t+2}, \ldots, \theta^{k}_{T-1})$. Then, at period $t$, the MMCC algorithm updates $\theta^{k-1}_{t}$ to be $\theta^{k}_{t}$ such that
{\allowdisplaybreaks
	\begin{align}\label{eq:monot}
	& U(c_0^{k-1}, \theta_1^{k-1}, \theta_2^{k-1}, \ldots, \theta_{t-1}^{k-1}, \theta_{t}^{k}, \theta_{t+1}^{k}, \ldots, \theta_{T-1}^{k})\notag\\
	\geq{} & U(c_0^{k-1}, \theta_1^{k-1}, \theta_2^{k-1}, \ldots, \theta_{t-1}^{k-1}, \theta_{t}^{k-1}, \theta_{t+1}^{k}, \ldots, \theta_{T-1}^{k}),
	\end{align}}
which can be easily shown to be equivalent to \eqref{equ:new_2};
see Appendix \ref{app:simple_deriv} for a detailed proof.
Therefore, such $\theta^{k}_{t}$ that satisfies \eqref{eq:monot} can be obtained by finding a suboptimal (optimal) solution to \eqref{eq:opt_t_S}.

Similarly, at period 0, before $c_0^{k-1}$ is updated, the control policy parameter is
$(c^{k-1}_0, \theta^{k}_{1}, \ldots, \theta^{k}_{T-1})$. Then, the MMCC algorithm updates $c_0^{k-1}$ to be $c_0^k$ such that
\begin{align}\label{eq:mono_0}
& U(c_0^{k}, \theta_1^{k}, \theta_2^{k}, \ldots, \theta_{T-1}^{k}) \geq U(c_0^{k-1}, \theta_1^{k}, \theta_2^{k}, \ldots, \theta_{T-1}^{k}).
\end{align}
Algorithm \ref{alg:Main_Algorithm} summarizes the MMCC algorithm for solving problem \eqref{equ:multi_per_obj}.

\begin{algorithm}[!htbp]
	\caption{The MMCC algorithm for solving problem \eqref{equ:multi_per_obj}.}
	\label{alg:Main_Algorithm}
	\begin{enumerate}
		\item Initialize $k=1$ and $x^0=(c^0_0, \theta^0_{1}, \theta^0_{2}, \ldots, \theta^0_{T-1})$.  
		\item Iterate $k$ until some stopping criteria are met. In the $k$th iteration, update $x^{k-1}=(c^{k-1}_0, \theta^{k-1}_{1}, \theta^{k-1}_{2}, \ldots, \theta^{k-1}_{T-1})$ to $x^{k}=(c^{k}_0, \theta^{k}_{1}, \theta^{k}_{2}, \ldots, \theta^{k}_{T-1})$
		by moving backwards from $t=T-1$ to $t=0$ as follows:
		\item[(a)] Move backward from $t=T-1$ to $t=1$. At each period $t$, update $\theta_{t}^{k-1}$ to be $\theta_{t}^{k}$ such that
		{\allowdisplaybreaks
			\begin{align} \label{equ:new_2}
			&  \mathbb{E}_0\left[\sum_{j=t}^{T-1} u_{j+1}(s_{j+1},s_{j},c_j)\middle | c_0^{k-1}, \theta_1^{k-1}, \ldots, \theta_{t-1}^{k-1}, \theta_t^k, \theta_{t+1}^k, \ldots, \theta_{T-1}^k \right]\notag\\
			\geq{} &  \mathbb{E}_0\left[\sum_{j=t}^{T-1} u_{j+1}(s_{j+1},s_{j},c_j)\middle | c_0^{k-1}, \theta_1^{k-1}, \ldots, \theta_{t-1}^{k-1}, \theta_t^{k-1}, \theta_{t+1}^k, \ldots, \theta_{T-1}^k \right],
			\end{align}}
		Such $\theta^{k}_{t}$ can be set as a suboptimal (optimal) solution to the problem
		\begin{align}\label{eq:opt_t_S}
		&    \max_{\theta_{t}\in \Theta_t}
		\mathbb{E}_0\left[\sum_{j=t}^{T-1} u_{j+1}(s_{j+1},s_j, c_j)\middle | c_0^{k-1}, \theta_1^{k-1}, \ldots, \theta_{t-1}^{k-1}, \theta_t, \theta_{t+1}^k, \ldots, \theta_{T-1}^k\right].
		\end{align}
  		where $\Theta_t=\{\theta\in \R^d\mid (c_0^{k-1}, \theta_1^{k-1}, \ldots, \theta_{t-1}^{k-1}, \theta, \theta_{t+1}^k, \ldots, \theta_{T-1}^k)\in\Theta\}$.

		\item[(b)] At period $0$, update $c_{0}^{k-1}$ to be $c_{0}^{k}$ such that
		\begin{align}
		    &\mathbb{E}_0\left[\sum_{j=0}^{T-1} u_{j+1}(s_{j+1}, s_j, c_j)\middle | c_0^k, \theta_1^{k}, \ldots, \theta_{T-1}^k \right]\notag
		\\\geq &\mathbb{E}_0\left[\sum_{j=0}^{T-1} u_{j+1}(s_{j+1}, s_j, c_j)\middle | c_0^{k-1}, \theta_1^{k}, \ldots, \theta_{T-1}^k \right].
		\end{align}
		Such $c^{k}_0$ can be set as a suboptimal (optimal) solution to the problem
		\begin{align}\label{eq:opt_0}
		&\max_{c_0\in \Theta_0}
		\mathbb{E}_0\left[\sum_{j=0}^{T-1} u_{j+1}(s_{j+1},s_j, c_j)\middle | c_0, \theta_1^{k}, \ldots, \theta_{T-1}^{k} \right],
		\end{align}
		where $\Theta_0=\{c\in\mathbb{R}^{n_c}\mid (c,\theta_1^{k}, \ldots, \theta_{T-1}^{k})\in\Theta \}$.
	\end{enumerate}
\end{algorithm}

Two remarks are in order: (i) In the MMCC algorithm when we update $\theta_t^{k-1}$ to $\theta_t^k$ or update $c_0^{k-1}$ to $c_0^k$ if no improvement of the objective function can be found, we simply set $\theta_t^k=\theta_t^{k-1}$ or set $c_0^{k}=c_0^{k-1}$.
(ii) Because the MMCC algorithm does not use the Bellman equation, it
can be applied to general control problems.

The intuition of the MMCC algorithm is also related to
the block coordinate descent (BCD) algorithms, in which the coordinates are divided into blocks and only one block of coordinates is updated at each sub-step of iterations in a cyclic order. However, the details of the two algorithms differ significantly:
(i) In essence, the MMCC algorithm attempts to update control policies in the control policy spaces (e.g.,  in the space of deep neural networks) rather than the Euclidean space. Consequently, all the parameters $\theta_t$ associated with the control policy $c_t$ are updated simultaneously, rather than block-wise one by one as in BCD. This is similar to the relationship between the classical EM (Expectation-Maximization) algorithm\footnote{See, e.g., \cite*{Dempster1977},  
	\citet*{Meng-Rubin-1993},
	and \citet[][Chap. 13]{Lange-2010}, among others. 
	For discussion on the connection between reinforcement learning and the EM algorithm, see \citet*{Dayan-Hinton-1997}.
} and BCD; in fact, \citet*{Neal-Hinton-1999} show that the EM algorithm can be viewed as a generalized BCD searching in the functional space of probability distribution functions rather than in the space of real numbers. 
(ii) BCD methods are used for maximizing deterministic objective functions, but the MMCC algorithm is used for maximizing the expectation of a random utility function (i.e., \eqref{equ:utility_func}), which generally cannot be evaluated analytically. That is why we have to employ simulation and stochastic optimization to implement the MMCC algorithm (see Section \ref{sec:Simulation_Based_Algorithm}).
(iii) The MMCC algorithm is more flexible in the optimization requirement.
Unlike the BCD algorithms, the MMCC algorithm does not update the control parameter based on the gradient of the objective function, mainly because in the problems solvable by the MMCC algorithm typically neither the objective function (i.e., \eqref{equ:utility_func}) nor the gradient of the objective function can be evaluated analytically.
(iv) The convergence of the MMCC algorithm holds under weaker conditions. Indeed, the convergence of the BCD algorithms is obtained based on various assumptions on the objective function such as that the objective function is convex or is the sum of a smooth function and a convex separable function or satisfies certain separability and regularity conditions;\footnote{See, e.g., \citet*{Luo-Tseng-1992}, \citet[][Chap. 3.7]{bertsekas2016nonlinear}, \citet*{Tseung-2001} and \citet*{Wright-2015}.} In contrast, the proof of convergence of the MMCC algorithm is similar to that of the EM algorithm, as in \cite*{wu1983convergence}, which does not need such assumptions on the objective function.  See Section \ref{sec:Convergence-Analysis} for details.
(v) Unlike some BCD algorithms, the MMCC algorithm 
does not require updating the control parameter to be the exact minimizer of the subproblem (\eqref{eq:opt_t_S} or \eqref{eq:opt_0}).
(vi) The setting of the MMCC algorithm is quite different from BCD. Indeed, the MMCC is implemented via deep neural network representation of policy functions, and the parameters can be updated using stochastic gradient descent.

\section{Convergence Analysis}
\label{sec:Convergence-Analysis}

The convergence properties of the MMCC algorithm are similar to those of the EM algorithm.
First, the MMCC algorithm has monotonicity in each iteration. Second, under mild assumptions, the sequence of objective function values generated by the iteration of the MMCC algorithm converges to a stationary value (i.e., objective function value evaluated at a stationary point) or a local maximum value. Third, the sequence of control parameters generated by the iteration of the MMCC algorithm converges to a stationary point or a local maximum point under some additional regularity conditions.

\subsection{Monotonicity}

\begin{theorem}\label{thm:monoto}
	The objective function $U(\cdot)$ defined in \eqref{equ:utility_func} monotonically increases in each iteration of the MMCC algorithm, i.e. for each $k \geq 1$,
	\be\label{equ:monoto}
	U(x^k)=U(c^k_0, \theta^k_1, \theta^k_2, \ldots, \theta^k_{T-1})\geq U(x^{k-1})=U(c^{k-1}_0, \theta^{k-1}_1, \theta^{k-1}_2, \ldots, \theta^{k-1}_{T-1}).
	\ee
\end{theorem}
\proof{}
See Appendix \ref{app:proof_monoto}.   
\endproof

\subsection{Convergence of the Value Function to a Stationary Value or a Local Maximum Value}
Let $\{x^k\}_{k\geq 0}$ be the sequence of control parameters generated by the MMCC algorithm. In this subsection, we consider the issue of the convergence of $U(x^k)$ to a stationary value or a local maximum value. We make the following mild assumptions on the objective function $U(\cdot)$ defined in \eqref{equ:utility_func}:
{\allowdisplaybreaks
	\begin{align}
	&\text{For any }x^0\ \text{such that}\ U(x^0)>-\infty, \{x\in\Theta\mid U(x)\geq U(x^0)\}\ \text{is compact.} \label{equ:assump_1}\\
	&U(\cdot)\ \text{is continuous in}\ \Theta\ \text{and differentiable in the interior of}\ \Theta.\label{equ:assump_2}
	\end{align}}
The assumption \eqref{equ:assump_2} is needed to define stationary points of $U(\cdot)$.

Suppose the objective function $U(\cdot)$ satisfies \eqref{equ:assump_1} and \eqref{equ:assump_2}. Then,
\be\label{equ:bounded_above}
\{U(x^k)\}_{k\geq 0}\ \text{is bounded above for any }x^0\ \text{such that}\ U(x^0)>-\infty.
\ee
By \eqref{equ:monoto} and \eqref{equ:bounded_above}, $U(x^k)$ converges monotonically to some $U^*$. However, it is not guaranteed that $U^*$ is a local maximum 
of $U$ on $\Theta$. Indeed, if the objective function $U$ has several local maxima and stationary points, which type of points the sequence generated by the MMCC algorithm converges to depends on the choice of the starting point $x^0$; this is also true in the case of the EM algorithm.

A map $\rho$ from points of $X$ to subsets of $X$ is called a point-to-set map on $X$ (\cite*{wu1983convergence}). Let $M$ be the point-to-set map of the MMCC algorithm defined in \eqref{equ:def_pts_map}. Define
{\allowdisplaybreaks
	\begin{align}
	\mathcal{M}&:=\text{set of local maxima of }U(\cdot)\ \text{in}\ \Theta,\notag\\
	\mathcal{S}&:=\text{set of stationary points of }U(\cdot)\ \text{in}\ \Theta,\notag\\
	\mathcal{M}(a)& :=\{x\in \M \mid U(x)=a\},\label{equ:M_inverse}\\
	\mathcal{S}(a)&:=\{x\in \cS \mid U(x)=a\}.\label{equ:S_inverse}
	\end{align}}
    
\begin{theorem}
	\label{thm:convergence_red_em}
	(Convergence of the value function). 
	Suppose the objective function $U$ satisfies conditions \eqref{equ:assump_1} and \eqref{equ:assump_2}. Let $\{x^k\}_{k\geq 0}$ be the sequence generated by $x^k\in M(x^{k-1})$ in the MMCC algorithm.
	
	(1) Suppose that
	\be\label{equ:cond_converg}
	U(x^k)>U(x^{k-1})\ \text{for any}\ x^{k-1}\notin \mathcal{S} ({\it \text{resp.}\ x^{k-1}\notin \mathcal{M}}).
	\ee
	Then, all the limit points of $\{x^k\}_{k\geq 0}$ are stationary points (resp. local maxima) of $U$, and $U(x^k)$ converges monotonically to $U^*=U(x^*)$ for some $x^*\in \mathcal{S}$ (resp. $x^*\in\mathcal{M}$).
	
	(2) Suppose that at each iteration $k$ in the MMCC algorithm and for all $t$,
	$\theta_t^{k}$ and $c_0^{k}$ are the optimal solutions to the problems 
	\eqref{eq:opt_t_S} and \eqref{eq:opt_0} respectively. Then, all the limit points of $\{x^k\}$ are stationary points of $U$ and $U(x^k)$ converges monotonically to $U^*=U(x^*)$ for some $x^*\in \mathcal{S}$.
\end{theorem}
\proof{}
See Appendix \ref{app:proof_red_em}.  
\endproof

\subsection{Convergence of the Control Policy to a Stationary Point or a Local Maximum Point}
Let $\M(a)$ and $\cS(a)$ be defined in \eqref{equ:M_inverse} and \eqref{equ:S_inverse}  respectively. Under the conditions of Theorem \ref{thm:convergence_red_em}, $U(x^k)\to U^*$ and all the limit points of $\{x^k\}$ are in $\cS(U^*)$ (resp. $\M(U^*)$). This does not imply the convergence of $\{x^k\}_{k\geq 0}$ to a point $x^*$. 
However, the following theorem provides sufficient conditions under which $x^k\to x^*$.

\begin{theorem}\label{thm:converg_x}
	(Convergence of the control policy). 
	Let $\{x^k\}_{k\geq 0}$ be an instance of an MMCC algorithm satisfying the conditions of Theorem \ref{thm:convergence_red_em}, and let $U^*$ be the limit of $\{U(x^k)\}_{k\geq 0}$.
	
	(1) If $\cS(U^*)=\{x^*\}$ (resp. $\M(U^*)=\{x^*\}$), then $x^k\to x^*$ as $k\to \infty$.
	
	(2) If $\|x^{k+1}-x^k\|\to 0$ as $k\to\infty$, then, all the limit points of $x^k$ are in a connected and compact subset of $\cS(U^*)$ (resp. $\M(U^*)$).
	In particular, if $\cS(U^*)$ (resp. $\M(U^*)$) is discrete, i.e., its only connected components are singletons, then $x^k$ converges to some $x^*$ in $\cS(U^*)$ (resp. $\M(U^*)$).
\end{theorem}
\proof{}
See Appendix \ref{app:proof_cong_x}.
\endproof

\section{An Implementation of the MMCC Algorithm}\label{sec:Simulation_Based_Algorithm}

\subsection{Implementing the MMCC Algorithm by Simulation}
In the MMCC algorithm, we need to find a suboptimal (optimal) solution to the problems 
\eqref{eq:opt_t_S} and \eqref{eq:opt_0}.
In practice, the expectation in the objective functions of these problems may not be evaluated analytically.
We propose solving these problems using stochastic gradient descent algorithms such as Adam for deep neural networks. At each iteration of the MMCC algorithm, sample paths are simulated using the current policy, and then a Monte Carlo optimization algorithm is applied to find updates of the control policy at each period to improve the objective function.

More precisely, at the beginning of the $k$th iteration, we first simulate $N$ i.i.d. (independently and identically distributed) sample paths of the states $(s_0, s_1, \ldots, s_{T-1})$ according to the control parameter $(c^{k-1}_0, \theta_1^{k-1}, \ldots, \theta_{T-1}^{k-1})$, which are obtained at the end of the $(k-1)$th iteration. We denote these sample paths as
$
(s_0, s_{1,l}^{k}, s_{2,l}^{k}, \ldots, s_{T-1,l}^{k}), l=1, \ldots, N.\notag
$

Furthermore, we divide these $N$ sample paths into $m$ minibatches, each containing $b$ sample paths. Let the sample paths in the $i$th minibatch be denoted as
$
(s_0, s_{1,l,i}^{k}, s_{2,l,i}^{k}, \ldots, s_{T-1,l,i}^{k}), l=1, \ldots, b, i=1, \ldots, m,
$ where $b\cdot m=N$.

In step 2(a) of Algorithm \ref{alg:Main_Algorithm}, for the $i$th minibatch, the expectation in the objective function of \eqref{eq:opt_t_S} is equal to
{\allowdisplaybreaks
	\begin{align}\label{equ:simu_E_t}
	&\mathbb{E}_0\left[\sum_{j=t}^{T-1} u_{j+1}(s_{j+1}, s_j, c_j)\middle | c_0^{k-1}, \theta_1^{k-1}, \ldots, \theta_{t-1}^{k-1}, \theta_t, \theta_{t+1}^k, \ldots, \theta_{T-1}^k\right]\notag\\
	= & \mathbb{E}_0\left\{\frac{1}{b}\sum_{l=1}^b \left[u_{t+1}(s^k_{t+1,l,i}(\theta_t), s^k_{t,l,i}, c^k_{t,l,i}(\theta_t))\phantom{\sum_{j=t+1}^{T-1}}\right.\right.\notag\\
	&\quad \quad \quad \quad \quad \quad  + \left.\left.\sum_{j=t+1}^{T-1} u_{j+1}(s^k_{j+1,l,i}(\theta_t), s^k_{j,l,i}(\theta_t), c^k_{j,l,i}(\theta_t))\right]\right\},
	\end{align}}
where $c^k_{t,l,i}(\theta_t)=c(t, s^k_{t,l,i}, \theta_{t})$ (see \eqref{equ:control}) and
$$(s^k_{t+1,l,i}(\theta_{t}), c^k_{t+1,l,i}(\theta_t), \ldots, s^k_{T-1,l,i}(\theta_{t}), c^k_{T-1,l,i}(\theta_t), s^k_{T,l,i}(\theta_{t}))$$
is a simulated sample path that starts from $s^k_{t,l,i}$ and then follows the control parameter $\theta_{t}, \theta_{t+1}^k, \ldots, \theta_{T-1}^k$. For each minibatch $i=1,\ldots, m$, the MMCC algorithm uses
{\allowdisplaybreaks
	\begin{align}\label{equ:appro_E_t}
	\tilde f(\theta_{t}):=& \frac{1}{b}\sum_{l=1}^b \left[u_{t+1}(s^k_{t+1,l,i}(\theta_t), s^k_{t,l,i}, c^k_{t,l,i}(\theta_t))\phantom{\sum_{j=t+1}^{T-1}}\right.\notag\\
	&\quad \quad \quad \quad \quad \quad  + \left.\sum_{j=t+1}^{T-1} u_{j+1}(s^k_{j+1,l,i}(\theta_t), s^k_{j,l,i}(\theta_t), c^k_{j,l,i}(\theta_t))\right]
	\end{align}}
as a realization of $\sum_{j=t}^{T-1} u_{j+1}(s_{j+1}, s_j, c_j)$ and applies Adam algorithm to update the parameter $\theta_t$ once.
Hence, at each iteration $k$ of the MMCC algorithm, in order to update the parameter $\theta_t$, we only need to simulate $N$ sample paths of the states during period $t+1$ to period $T$, i.e., $
(s_{t+1,l,i}^{k}(\theta_t), s_{t+2,l,i}^{k}(\theta_t), \ldots, s_{T,l,i}^{k}(\theta_t)), l=1, \ldots, b, i=1, \ldots, m,
$ where $b\cdot m=N$.

Similarly, in step 2(b)  of Algorithm \ref{alg:Main_Algorithm}, for the $i$th minibatch, the expectation in \eqref{eq:opt_0} is equal to
{\allowdisplaybreaks
	\begin{align}\label{eq:simu_E_0}
	& \mathbb{E}_0\left[\sum_{j=0}^{T-1} u_{j+1}(s_{j+1}, s_j, c_j)\middle | c_0, \theta_1^{k}, \ldots, \theta_{T-1}^k \right]\notag\\
	={} & \mathbb{E}_0\left\{\frac{1}{b}\sum_{l=1}^b\left[u_{1}(s^k_{1,l,i}(c_0), s_0, c_0) +\sum_{j=1}^{T-1} u_{j+1}(s^k_{j+1,l,i}(c_0), s^k_{j,l,i}(c_0), c^k_{j,l,i}(c_0))\right]\right\},
	\end{align}}
where $\left\{s_{1,l,i}^{k}\left( c_0 \right),c_{1,l,i}^{k}\left( c_0 \right),\ldots,s_{T-1,l}^{k}\left( c_0 \right), c_{T-1,l,i}^{k}\left(c_0 \right), s_{T,l,i}^{k}\left( c_0 \right)\right\}_{l=1}^{b}$ are $b$ i.i.d. sample paths of $(s_{1}, c_{1}, \ldots, s_{T-1}, c_{T-1}, s_T)$ that are simulated starting from $s_0$ and then following the control parameters $(c_0, \theta_1^{k}, \ldots, \theta_{T-1}^k)$. The MMCC algorithm uses
\begin{align}\label{equ:appro_E_0}
\tilde f(c_0):=\frac{1}{b}\sum_{l=1}^b \left[u_{1}(s^k_{1,l,i}(c_0), s_0, c_0) +\sum_{j=1}^{T-1} u_{j+1}(s^k_{j+1,l,i}(c_0), s^k_{j,l,i}(c_0), c^k_{j,l,i}(c_0))\right]
\end{align}
as a realization of $\sum_{j=0}^{T-1} u_{j+1}(s_{j+1}, s_j, c_j)$ based on the $i$th minibatch when solving the problem \eqref{eq:opt_0}.

At each iteration $k$ and for each time step $t$ in the algorithm, when the Adam algorithm is used for maximizing \eqref{equ:simu_E_t} and \eqref{eq:simu_E_0}, we update the parameters $m$ times by using $m$ minibatches of sample. Then, the computational cost of solving the problems 
\eqref{eq:opt_t_S} and \eqref{eq:opt_0} are respectively 
$O(m(T-t+1))$ and $O(mT)$. Hence, the computational cost of each iteration of the MMCC algorithm is $O(mT^2)$.

\subsection{A Numerical Example: A Forward-Backward Stochastic Differential Equation}
\label{sec:BSDEExample}

There are intrinsic links between recursive utilities and forward-backward stochastic differential equations (FBSDEs); in addition, there are also connections between three mathematical concepts, namely FBSDEs, stochastic control, and semi-linear PDEs. See, e.g., \citet*{Peng1997}, \citet*{ShroderSkiadas}, \citet*{Pham2015} and \citet*{Pham2015b}.

In this subsection, we shall first
give a brief and nontechnical outline of the key connections between the
three, which will be used later when we numerically solve portfolio choice
for recursive utilities under stochastic volatility. Then we shall give an example where an FBSDE with 100 dimensions can be solved analytically so that we have a benchmark to demonstrate the effectiveness of the MMCC algorithm.

Consider a stochastic control problem
\begin{equation}\label{equ:fbsde_obj}
\min_{y,Z(\cdot )}\mathbb{E}\left\vert Y_{T}^{y,X,Z}-g(X_{T})\right\vert ^{2}, 
\end{equation}
where the dynamics of $Y$ and $X$ are given by
\begin{eqnarray}
dY_{t}^{y,X,Z} &=& -f(t,X_{t},Y_{t},Z_{t})dt+Z_{t}^{\top }dW_{t},~~~
Y_{0}^{y,X,Z}=y,  \label{e1} \\
dX_{t} &=& \mu (t,X_{t})dt+\sigma (t,X_{t})^{\top }dW_{t},~~~X_{0}=x_0,
\label{e3}
\end{eqnarray}
where $\top $ means transpose and $W_{t}$ is a $d_w$-dimensional Brownian
motion. Note that this is a non-standard control problem, as the initial
value $y$ is also a control variable. It can be shown that under mild
conditions, the value of the control problem is zero; in fact, this control
has a natural interpretation in option pricing, $y$ is the initial option
price and the $Z$ is the hedging strategy. Under mild conditions, the solution is given by
\[
Y_{t}=g(X_{T})+\int_{t}^{T}f(s,X_{s},Y_{s},Z_{s})ds-\int_{t}^{T}Z_{s}^{\top
}dW_{s}. 
\]
The optimal control ($Y_{t}^{\ast },\ Z_{t}^{\ast }$) in \eqref{equ:fbsde_obj} is
linked to a semi-linear PDE
\begin{eqnarray}
&&\frac{\partial u}{\partial t}(t,x)+\frac{1}{2}\mathrm{Tr}\left[ \sigma(t, x)
\sigma(t,x)^{\top}\mathrm{Hess}_{x}u(t,x)\right] +\nabla_x u(t,x)^{\top} \mu
(t,x)  \label{e2} \\
\ &&\ \ +f(t,x,u(t,x),\sigma (t,x)^{\top }\nabla_x u(t,x))=0,  t\in[0, T)\nonumber
\end{eqnarray}
with the terminal condition
$
u(T,x)=g(x), 
$
via 
\[
y^*=u(0, x_0),\ Y_{t}^{\ast }=u(t,X_{t}),\ \ Z_{t}^{\ast }=\sigma (t,X_{t})^{\top }\nabla_x
u(t,X_t), 
\]
where $\mathrm{Tr}$, $\mathrm{Hess}$, and $\nabla $ means the trace, Hessian
matrix, and gradient operator, respectively. The optimal objective function value is 0. 

We discretize $[0, T]$ into $0=t_0<t_1<t_2<\cdots < t_{N_T}=T$. The discretized control problem is
\begin{equation*}
\min_{y,z_0,\theta_{t_1},\ldots,\theta_{t_{N_T-1}}}\mathbb{E}\left\vert Y_{T}^{y,X,Z}-g(X_{T})\right\vert ^{2}, 
\end{equation*}
subject to the dynamics
\begin{align}
X_{t_0}&=x_0, Y_{t_0}=y,Z_{t_0}=z_0,\notag\\
Z_{t_n}&=\sigma(t_n, X_{t_n})^\top c(t_n, X_{t_n}, \theta_{t_n}),\ 1\leq n \leq N_T-1,\notag\\
X_{t_{n+1}} &= X_{t_{n}}+\mu (t_n,X_{t_n})(t_{n+1}-t_n)+\sigma (t_n,X_{t_n})^\top (W_{t_{n+1}}-W_{t_n}),\ 0\leq n \leq N_T-1,\notag\\
Y_{t_{n+1}} &= Y_{t_{n}} - f(t,X_{t_n},Y_{t_n},Z_{t_n})(t_{n+1}-t_n)+Z_{t_n}^\top (W_{t_{n+1}}-W_{t_n}),\ 0\leq n \leq N_T-1. \notag
\end{align}
\item
where $z_{0}$ approximates $\sigma (0,x_{0})^{\top }\nabla_x
u(0,x_0)$; $c(t_n, x, \theta_{t_n})$ is a neural network approximation to the gradient function $\nabla_x u(t_n,x)$, and $\theta_{t_n}$ is the parameter of the network.

Now take a special case in the above definition with $\sigma =\sqrt{2}I$ where $I$ is the identity matrix, $\mu =\textbf{0}_{d_w\times 1}$ where $\textbf{0}_{d_w \times 1}$ is a $d_w$-dimensional vector with all entries equal to 0, then
 $X_{t}=x_0+\sqrt{2} W_t$ in \eqref{e3}.
With a particular choice
$
f(t,x,y,z)=\beta z^{\top}z$,
\eqref{e2} reduces to 
\[
\frac{\partial u}{\partial t}(t,x)+\Delta _{x}u(t,x)+\beta
\nabla_x u(t,x)^{\top}\nabla_x u(t,x)=0.
\]
The stochastic control in \eqref{equ:fbsde_obj} now becomes
\begin{align}
\min_{y,Z(\cdot )}  &\ \mathbb{E}\left\vert Y^{y,Z}_T-g(X_{T})\right\vert ^{2},\label{equ:fbsde_control_prob} \\
\text{s.t.} &\ dY^{y,Z}_t=-\beta Z_{t}^{\top}Z_t dt+Z_t^{\top}dW_t,\ \ \ Y_{0}^{y,Z}=y,\\
& \ dX_t=\sqrt{2}dW_t, \ \ \ X_0 = x_0.
\end{align}
In particular, the It\^{o}'s formula implies that the solution of the
semi-linear PDE is given by 
\[
u(t,x)=\frac{1}{\beta }\ln \left( \mathbb{E}\left[ \exp \left( \beta
g(x+W_{T-t})\right)\right] \right), 
\]
and the optimal $Y^*$ is
\[
Y_{t}^{\ast }=u(t,X_{t}) ,\ \ y^{\ast }=u(0,x_0)=\frac{1}{\beta }\ln \left(\mathbb{E}\left[ \exp \left(\beta g(x_0+W_T)\right) \right] \right);
\]
see \citet*{chass2016} and \citet*{EHan2017_published} for details.

In the numerical examples, we choose $d_w=100$, $T=1$, $\beta=-1$, $x_0= \textbf{0}_{d_w\times 1}$, and $g(x)=\ln\left(\frac{1}{2}\left(1+x^\top x\right)\right)$. We discretize the time interval $[0, T]$ into $N_T=20$ equal subintervals with ending points denoted as $t_{0}=0 < t_1 < \cdots < t_{N_T - 1} < t_{N_T}=T$. The control policy at time $t_n$ in the discretized problem is 
$c(t_n, x, \theta_{t_n})$. For each $t_n, n=1, \ldots, N_T-1$, we use a feed-forward neural network with parameter $\theta_{t_n}$ to approximate the control policy function. The neural network has six layers, where the input and output layers have 100 neurons and each of the four hidden layers has 110, 120, 120, and 110 neurons respectively. The nonlinear activation function of each layer is the rectified linear function. The minibatch size used in optimizing the neural network parameters is 64.

\begin{figure}[htb]
	\begin{centering}
		\includegraphics[width=\textwidth]{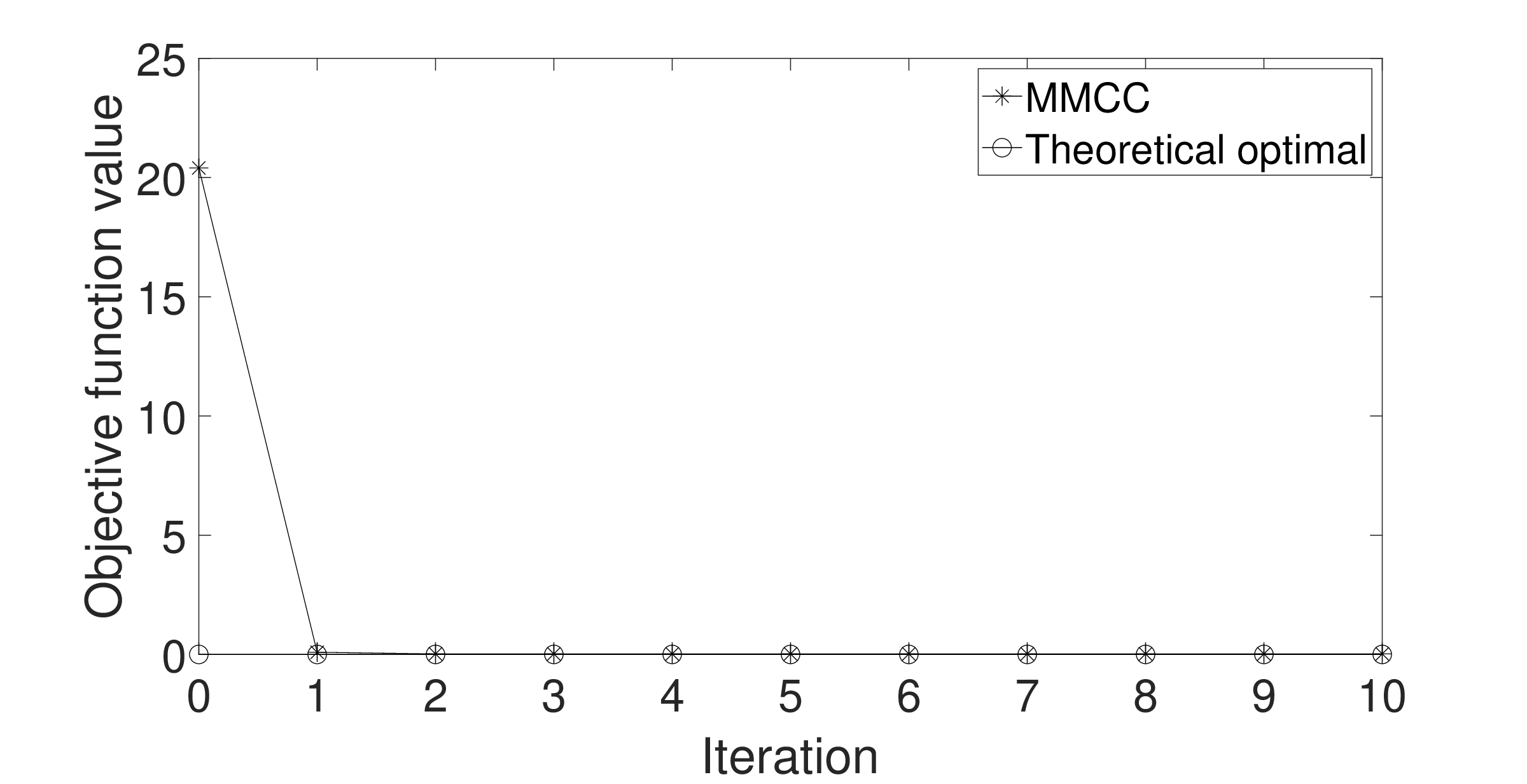}
	\end{centering}
	\caption{Objective function values of the MMCC algorithm defined in \eqref{equ:fbsde_obj}. The MMCC algorithm converges after 3 iterations. It uses $N=12,800$ sample paths in the simulation and $m=200$ iterations in the Adam algorithm. The initial learning rate of the Adam algorithm is set to be 0.01. It takes 1.5 hours for each iteration under a Python implementation of the MMCC algorithm based on TensorFlow. The optimal objective value obtained by the MMCC algorithm is 0.0229 (with a standard error of 0.0313). The standard error is equal to the sample standard deviation of the $N$ samples of the objective function in \eqref{equ:fbsde_obj} divided by $\sqrt{N}$. The theoretical optimal objective function value is 0. \label{fig:fbsde_opt_value_iter}}
\end{figure}

\begin{figure}[htb]
	\begin{centering}
		\includegraphics[width=\textwidth]{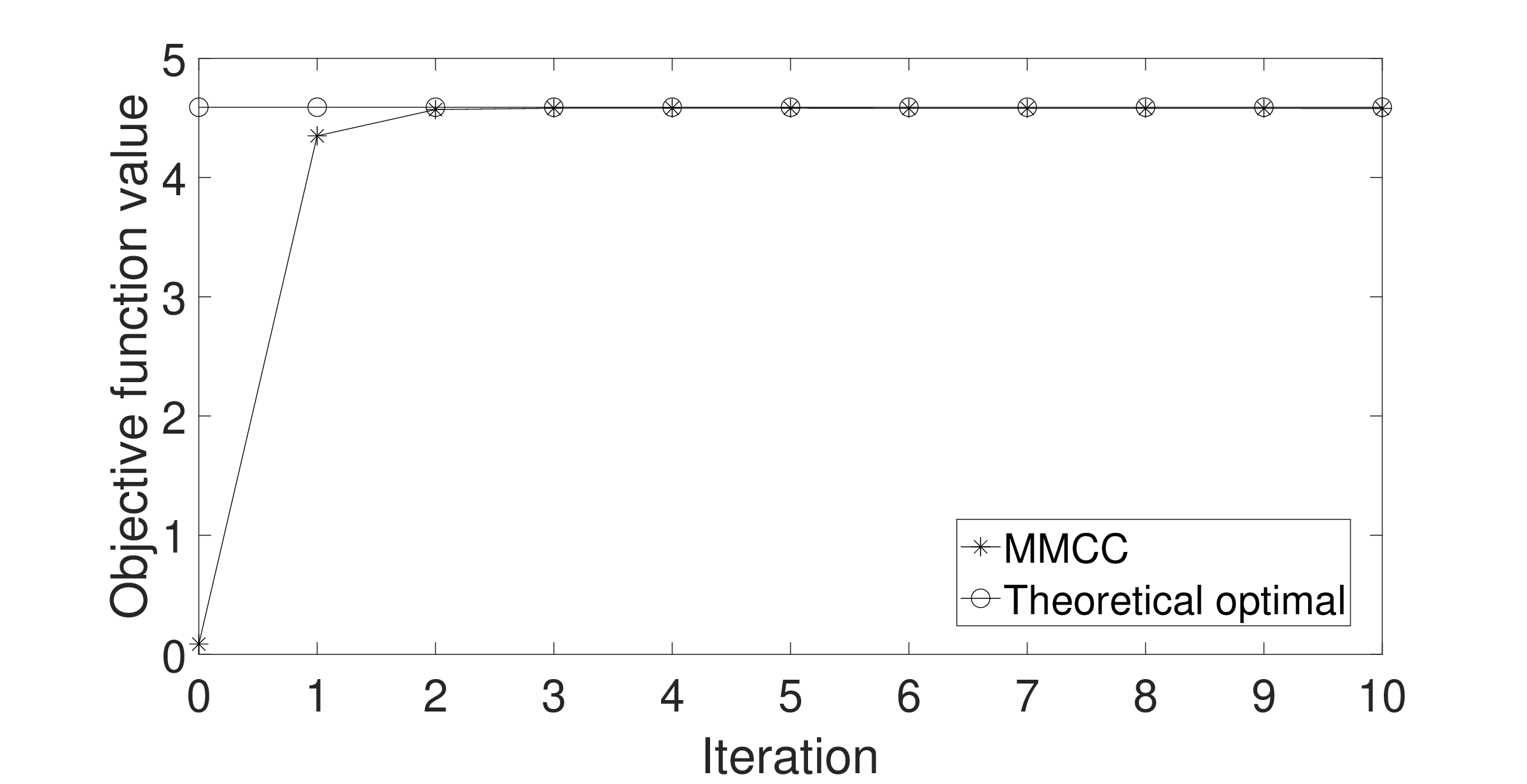}
	\end{centering}
	\caption{The value of $y^*$ in the control problem \eqref{equ:fbsde_control_prob} of the MMCC algorithm. The MMCC algorithm converges after 3 iterations. It uses $N=12,800$ sample paths in the simulation and $m=200$ iterations in the Adam algorithm. The initial learning rate of the Adam algorithm is set to be 0.01. It takes 1.5 hours for each iteration under a Python implementation of the MMCC algorithm based on TensorFlow. The optimal value $y^*$ obtained by the MMCC algorithm is 4.5799. The theoretically optimal value $y^*$ is 4.5901. \label{fig:fbsde_y_iter}}
\end{figure}

Figure \ref{fig:fbsde_opt_value_iter} shows the objective function values of the MMCC algorithm defined in \eqref{equ:fbsde_obj}. The MMCC algorithm converges after 3 iterations. It uses $N=12,800$ sample paths in the simulation and $m=200$ iterations in the Adam algorithm. The initial learning rate of the Adam algorithm is set to be 0.01. It takes about 1.5 hours for each iteration under a Python implementation of the MMCC algorithm based on TensorFlow. The optimal objective value obtained by the MMCC algorithm is 0.0229 (with a standard error of 0.0313). The standard error is equal to the sample standard deviation of the $N$ samples of the objective function in \eqref{equ:fbsde_obj} divided by $\sqrt{N}$. The theoretical optimal objective function value is 0. Figure \ref{fig:fbsde_y_iter} shows the value of $y^*$ in the control problem \eqref{equ:fbsde_control_prob} of the MMCC algorithm. The optimal value $y^*$ obtained by the MMCC algorithm is $4.5799$. The theoretically optimal value $y^*$ is 4.5901.

\section{Application 1: Recursive Utility with Stochastic Volatility}
\label{sec:recursiveUtility}

In this section, we consider maximizing the recursive utility in \citet*{EpsteinZin} under 
stochastic volatility models. Consider a market with two assets, a money
market account $M_{t}$ with a fixed risk-free rate $r$,
\[
dM_{t}=rM_{t}dt, 
\]
and a stock $S_{t}$ with a stochastic volatility modeled by
\begin{eqnarray*}
	\frac{dS_{t}}{S_{t}} &=&\left(r+\lambda (Y_{t})\right) dt+\sigma (Y_{t})dW_{t},\\
	dY_{t} &=&\alpha (Y_{t})dt+\beta (Y_{t})\left[ \rho W_{t}+\sqrt{1-\rho^2}
	\tilde{W}_{t}\right], Y_0=y_0,
\end{eqnarray*}
where ${W}_t$ and $\tilde{W}_t$ are two independent standard Brownian motions.
In the Heston model \citep{Heston},
\[
\alpha (y)=\theta -\kappa y,\ \ \beta (y)=\bar{\beta}\sqrt{y},\ \ \sigma (y)=
\sqrt{y};\ \ \lambda (y)=\bar{\lambda}y,\ \ \bar{\beta},\bar{\lambda}>0;\ \ 
\]
In the inverse Heston model \citep*{ChackoViceira},
\[
\alpha (y)=\theta -\kappa y,\ \ \beta (y)=\bar{\beta}\sqrt{y},\ \ \sigma (y)=
\frac{1}{\sqrt{y}};\ \ \lambda (y)=\bar{\lambda},\ \ \bar{\beta},\bar{\lambda
}>0. 
\]

Suppose $\pi _{t}$ is the proportion of the investor's wealth invested in the stock at time $t$, and $c_t$ is the consumption rate at time $t$. Then, the dynamics of total wealth $X_t$ of the investor are given by
\[
dX_{t}=X_{t}\left[ \left( r+\pi _{t}\lambda (Y_{t})\right) dt+\pi _{t}\sigma
(Y_{t})dW_{t}\right] -c_{t}dt.
\]

Consider a recursive utility $V_{t}$ defined as
\[
V_{t}=\mathbb{E}\left[\left. \int_{t}^{T}f(c_{s},V_{s})ds+U(X_{T})\right|\mathcal{F}_{t}\right]
,\ \text{where}\ U(x)=\frac{1}{1-\gamma }x^{1-\gamma }. 
\]
Note that the Epstein-Zin recursive utility is given by, when $\psi >0$, $
\psi \neq 1$,
\[
f(c,v)=\delta \theta v\left[ \left( \frac{c}{\left( (1-\gamma )v\right)
	^{1/(1-\gamma )}}\right) ^{1-\frac{1}{\psi }}-1\right],\ \ \psi >0,\ \psi
\neq 1, 
\]
\[
\delta >0,\ \gamma >0,\ \gamma \neq 1;\ \theta =\frac{1-\gamma }{1-\frac{1}{\psi}}, 
\]
and when $\psi =1$,
\[
f(c,v)=\delta (1-\gamma )v\left[ \ln (c)-\frac{1}{1-\gamma }\ln \left(
(1-\gamma )v\right) \right],\ \ \ \psi =1. 
\]
\citet*{ChackoViceira} derives an analytical solution for the case $\psi
=1$ for infinite horizon ($T=\infty $) under the inverse Heston model. \citet*{Kraft2013}
solves the case 
\begin{equation}
\psi =2-\gamma +\frac{(1-\gamma )^{2}}{\gamma }\rho ^{2}  \label{en1}
\end{equation}
with finite $T$ under both the Heston model and the inverse Heston model\footnote{
	Some numerical methods based on a fixed point iteration are discussed in \citet*{Kraft2017}, though neither the Heston model nor the inverse
	Heston model is covered as their conditions (A1) and (A2) fail to hold.}.

In this section, we shall solve the problem completely for arbitrary $\psi >0
$, under both the Heston and inverse Heston model, using the MMCC algorithm
by using a connection between an FBSDE and a semi-linear PDE associated with
the recursive utility. Indeed, by the HJB equation, it can be shown (see
equations (4.3)-(4.5) in \citet*{Kraft2013}) that the value function is given by
\[
w(t,x,y)=\frac{1}{1-\gamma }x^{1-\gamma }g(t,y)^{k},\ \ k=\frac{\gamma }{
	\gamma +(1-\gamma )\rho ^{2}},
\]
and the optimal control policies are given by
\begin{align*}
	\pi ^{\ast }(t,y) &=\frac{1}{\gamma }\frac{\lambda (y)}{(\sigma (y))^{2}}+
	\frac{k}{\gamma }\frac{\beta (y)\rho }{\sigma (y)g(t,y)}\frac{\partial g(t,y)
	}{\partial y}, \\
	\frac{c^{\ast }(t,x,y)}{x} &=\delta ^{\psi }g(t,y)^{-\psi k/\theta }.
\end{align*}
Note that when $\psi =1$ and $\psi k/\theta =0$, then $\frac{c^{\ast }(t,x,y)}{x}
=\delta $. Here $g(t,y)$ solves a semi-linear PDE
\[
0=\frac{\partial g}{\partial t}-\tilde{r}(y)g+\tilde{\alpha}(y)\frac{
	\partial g}{\partial y}+\frac{1}{2}\beta ^{2}(y)\frac{\partial ^{2}g}{
	\partial y^{2}}+\frac{\theta \delta ^{\psi }}{\psi k}g^{1-\psi k/\theta }
\]
with the terminal condition $g(T,y)=1$, where
\begin{align*}
	\tilde{r}(y) &=\frac{1}{k}\left[ r(1-\gamma )-\delta \theta +\frac{1}{2}
	\frac{1-\gamma }{\gamma }\frac{\lambda ^{2}(y)}{\sigma ^{2}(y)}\right], \\
	\tilde{\alpha}(y) &=\alpha (y)+\frac{1-\gamma }{\gamma }\frac{\lambda (y)}{
		\sigma (y)}\beta (y)\rho .
\end{align*}

By the connection with FBSDE discussed in Section \ref{sec:BSDEExample}, we know that we need
to solve a stochastic control problem
\[
\min_{\xi _{0},Z(\cdot )}\mathbb{E}\left[ \left( \xi _{T}-1\right) ^{2}\right],
\]
subject to the dynamics
\begin{align*}
	d\xi _{t} &=\left[ \tilde{r}(\eta _{t})\xi _{t}-\frac{\theta \delta ^{\psi }}{\psi k}(\xi _{t})^{1-\psi k/\theta }\right] dt+Z_{t}dW_{t}, \\
	d\eta _{t} &=\tilde{\alpha}(\eta _{t})dt+\beta (\eta _{t})dW_{t},\ \ \ \eta
	_{0}=y_0,
\end{align*}
where $W_{t}$ is a standard one-dimensional Brownian motion starting from 0. 
Then 
\[
\xi _{0}=g(0,y_0). 
\]

\begin{figure}
\begin{centering}
	\includegraphics[width=\textwidth]{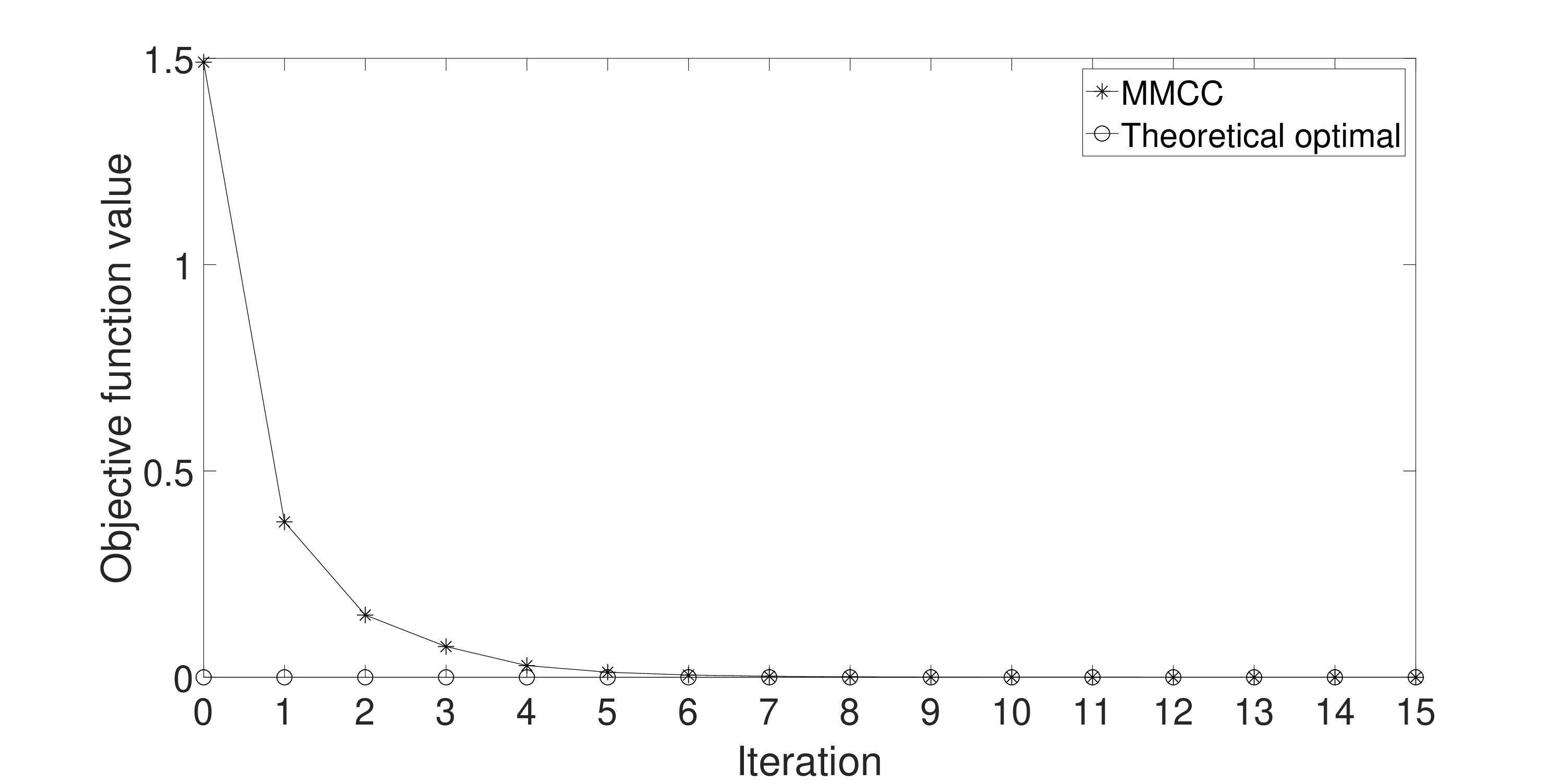}
\end{centering}
\caption{Objective function values of the MMCC algorithm defined in \eqref{equ:fbsde_obj}. The MMCC algorithm converges after 8 iterations. It uses $N=102,400$ sample paths in the simulation and $m=200$ iterations in the Adam algorithm. The initial learning rate of the Adam algorithm is set to be 0.01. It takes 36 minutes for each iteration under a Python implementation of the MMCC algorithm based on TensorFlow. The optimal objective value obtained by the MMCC algorithm is 2.4145e-6 (with a standard error of 1.4150e-7). The standard error is equal to the sample standard deviation of the $N$ samples of the objective function in \eqref{equ:fbsde_obj} divided by $\sqrt{N}$. The theoretical optimal objective function value is 0. \label{fig:stoc_vol_opt_value_iter}}
\end{figure}

\begin{figure}[htb]
\begin{centering}
	\includegraphics[width=\textwidth]{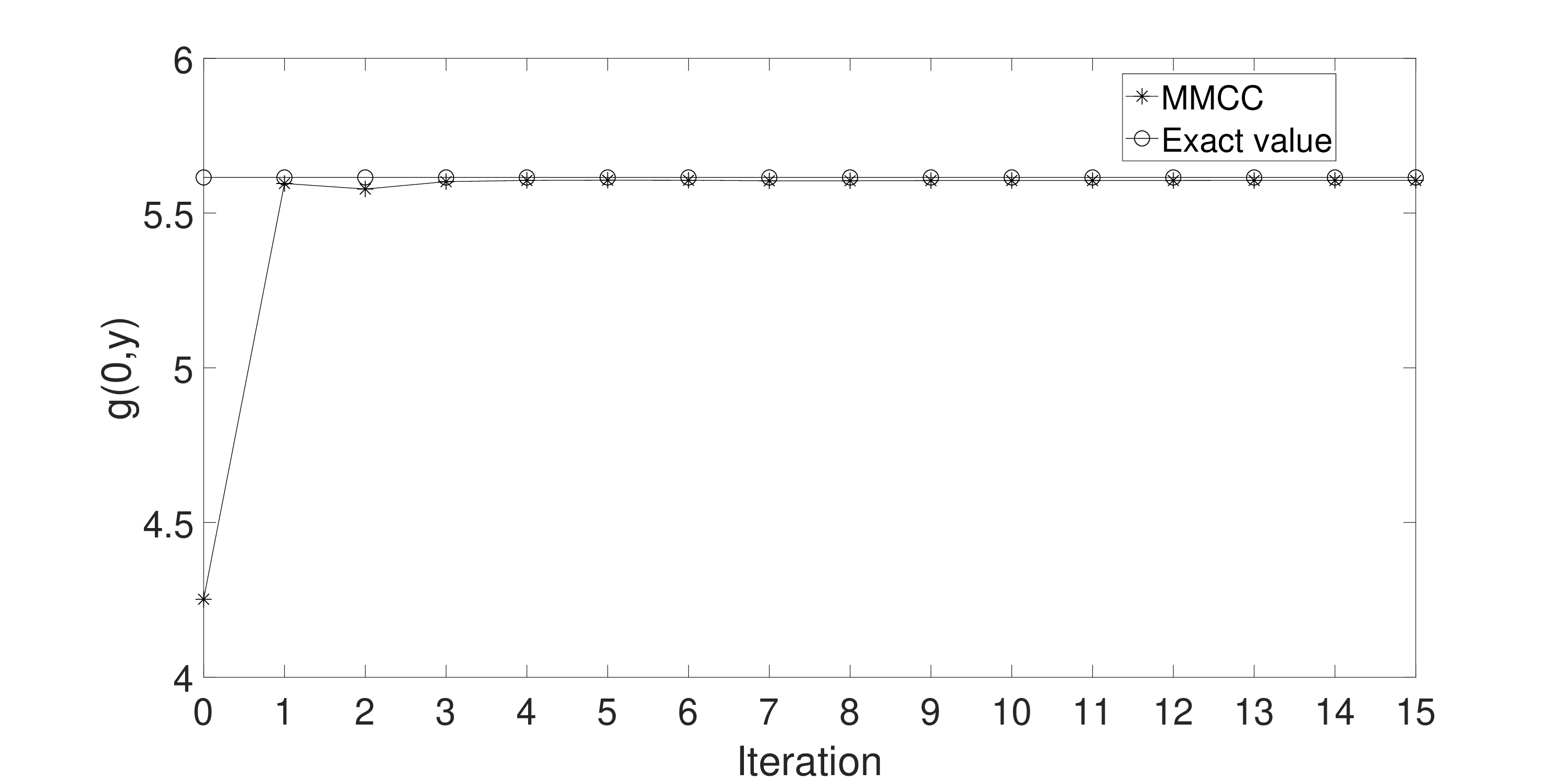}
\end{centering}
	\caption{The value of $g(0,y)$ for $y=0.1$ in the iteration of the MMCC algorithm. The MMCC algorithm converges after 8 iterations. It uses $N=102,400$ sample paths in the simulation and $m=200$ iterations in the Adam algorithm. The initial learning rate of the Adam algorithm is set to be 0.01. It takes 36 minutes for each iteration under a Python implementation of the MMCC algorithm based on TensorFlow. The value $g(0,y)$ obtained by the MMCC algorithm is $5.6054$. The exact value of $g(0,y)$ for $y=0.1$ is 5.6150. \label{fig:stoc_vol_y_iter}}
\end{figure}

We solve Heston's model for $T=10$ and discretize $[0, T]$ into 120 time periods. In the numerical example, we choose $r=0.05$, $\delta=0.08$, $\gamma=2$, $\rho = -0.5$, $\kappa=5$, $\Bar{y}=0.0225$, $\Bar{\lambda}=0.07/\sqrt{\Bar{y}}$, $\Bar{\beta}=0.25$. The neural network for $c(t_n, x, \theta_{t_n})$ has four layers, where the input layer and the output layer have 1 neuron and the two hidden layers have 120 neurons. 
The nonlinear activation function of each layer is the rectified linear function and the minibatch size in the Adam algorithm is 512.
We use $N=102,400$ sample paths in the simulation and $m=200$ iterations in the Adam algorithm.
The MMCC algorithm converged after 8 iterations. It takes 36 minutes for each iteration. Figure \ref{fig:stoc_vol_opt_value_iter} shows the objective function value in the iteration. The optimal objective value obtained by the MMCC algorithm is 2.4145e-6 (with a standard error of 1.4150e-7). The theoretical optimal objective function value is 0.
Figure \ref{fig:stoc_vol_y_iter} shows the value of $g(0,y_0)$ for $y_0=0.1$ in the iteration. The value $g(0,y_0)$ obtained by the MMCC algorithm is $5.6054$. The exact value of $g(0,y_0)$ for $y_0=0.1$ is 5.6150.

\section{Application 2: Multi-Sector Stochastic Growth}
\label{sec:business_cycle}

\subsection{Problem Formulation}

Starting from  \citet*{BrockM1972}, stochastic growth models play a fundamental role in macroeconomics, especially in    
the models related to real business cycle
(see e.g., \cite*{kydland1982time}, \cite*{long1983real}).
In the literature, this is typically studied assuming an infinite time horizon with one sector economy,
under which a stationary solution can be computed. In particular,
a log-linear linear-quadratic (LQ) approximation (see, e.g., \cite*{christiano1990linear}) is used to approximate
the objective function, which transforms the problem into a well-studied
linear-quadratic programming problem.

By using the MMCC algorithm, we can solve a more general multi-sector stochastic growth model with a finite time horizon, constant relative risk aversion (CRRA) utility, and general capital depreciation. 
There are significant differences between
the finite time horizon and infinite time horizon problem.

Consider a multi-sector model with $n$ commodities. The control variables
are $L_{it}$, the labor time input allocated to the production of commodity $
i$, and $X_{ijt}$, the quantity of commodity $j$ allocated for the
production of commodity $i$. The objective is 
\begin{equation}\label{equ:rbc_finite_obj}
\max_{L_{it}, Z_t, c_{jt}, X_{ijt}}\mathbb{E}_{0}\left[ \sum_{t=0}^{T}\beta ^{t}\left(
\sum_{i=1}^{n}\theta _{i}\frac{c_{it}^{1-\tau _{i}}}{1-\tau _{i}}+\theta _{0}
\frac{Z_{t}^{1-\tau _{0}}}{1-\tau _{0}}\right) \right],
\end{equation}
where $c_{it}$ is the consumption of the $i$th commodity and $Z_{t}$ is the
amount of leisure time consumed, subject to the state dynamics
\begin{align}
& Z_{t}+\sum_{i=1}^{n}L_{it}=H,\ t=0,1,2,\ldots, \label{equ:rbc_constr_1}\\
& c_{jt}+\sum_{i=1}^{n}X_{ijt}=Y_{jt},\ \ j=1,2,\ldots,n,\ t=0,1,2,\ldots, \label{equ:rbc_constr_2}\\
& Y_{i,t+1}=\lambda_{i,t+1}L_{it}^{b_{i}}\prod\limits_{j=1}^{n}X_{ijt}^{a_{ij}},\ i=1,2,\ldots,n,\label{equ:RBC_dynamics}
\end{align}
where the strictly positive stochastic noise $\lambda_{i,t+1}$ is an observable time
homogeneous Markov process; $H$, $b_i$, and $a_{ij}$ are given strictly positive constants; $b_{i}+\sum_{j=1}^{n}a_{ij}=1$.

When $T=\infty$ and $\tau _{i}=1$, $i=1,\ldots, n$ (i.e. all the utility functions are
logarithm), the model becomes the model in \cite*{long1983real}.
In this case, the optimal feedback control policies are given by
\begin{align}
L_{it}^{\ast } & =\frac{\beta \gamma _{i}b_{i}}{\theta _{0}+\beta
	\sum_{j=1}^{N}\gamma _{j}b_{j}}H,\ \ \ \ Z_{t}^{\ast
}=\frac{\theta _{0}}{\theta _{0}+\beta \sum_{j=1}^{N}\gamma _{j}b_{j}}H,\label{equ:rbc_opt_infinite_1}\\
X_{ijt}^{\ast }&=\frac{\beta \gamma _{i}a_{ij}}{\gamma _{j}}Y_{jt}^{\ast },\ \ \ \ c_{it}^{\ast }=\frac{\theta _{i}}{\gamma _{i}}Y_{it}^{\ast},\label{equ:rbc_opt_infinite_2}
\end{align}
where 
\[
\gamma_{j}=\theta_{j}+\beta \sum_{i=1}^{N}\gamma _{i}a_{ij},\ \text{or}\ \gamma ^{\top }=\theta ^{\top }(I-\beta A)^{-1},\ A=(a_{ij}).
\]

\subsection{Numerical Results}

We will compare the objective function value obtained by the MMCC algorithm with that obtained by the optimal policy for the infinite horizon problem given in \eqref{equ:rbc_opt_infinite_1} and \eqref{equ:rbc_opt_infinite_2}. The control policy at period $t$ is $p_t=(Z_t, (L_{it}), (c_{jt}), (X_{ijt}))$ which has a dimension of $(n+1)^2$, $t=0, 1, \ldots, T-1$. The state variable at time $t$ is $S_t=(Y_t, \lambda_t)$ with dimension $2n$. For each $t=1, \ldots, T-1$, we use a feed-forward neural network with parameter $\theta_t$ to approximate the control policy function. The neural network has four layers, where the input layer and the output layer have $2n$ and $(n+1)^2$ neurons, respectively. The two hidden layers have $300$ neurons. The nonlinear activation function of the first two layers is the rectified linear function, and the activation function of the output layer is a linear combination of $n+1$ softmax functions, which is used to impose the constraints in \eqref{equ:rbc_constr_1} and \eqref{equ:rbc_constr_2}. 

As in the numerical example of \cite*{long1983real}, we choose $n=6$ and $A$ as defined in \cite*{long1983real}. The parameters in the model dynamics are specified as $H=1.0$, $\beta=0.95$, $\theta=(\theta_0,\theta_1,\theta_2,\theta_3,\theta_4,\theta_5,\theta_6)=(0.1, 0.1, 0.12, 0.08, 0.1, 0.2, 0.3)$, $Y_0=(6,10,9,5,8,4)$, and $\log \lambda_{i,t+1}$ is i.i.d. as a standard normal distribution across $i$ and $t$. 

\begin{figure}[htb]
	\begin{centering}
		\includegraphics[width=\textwidth]{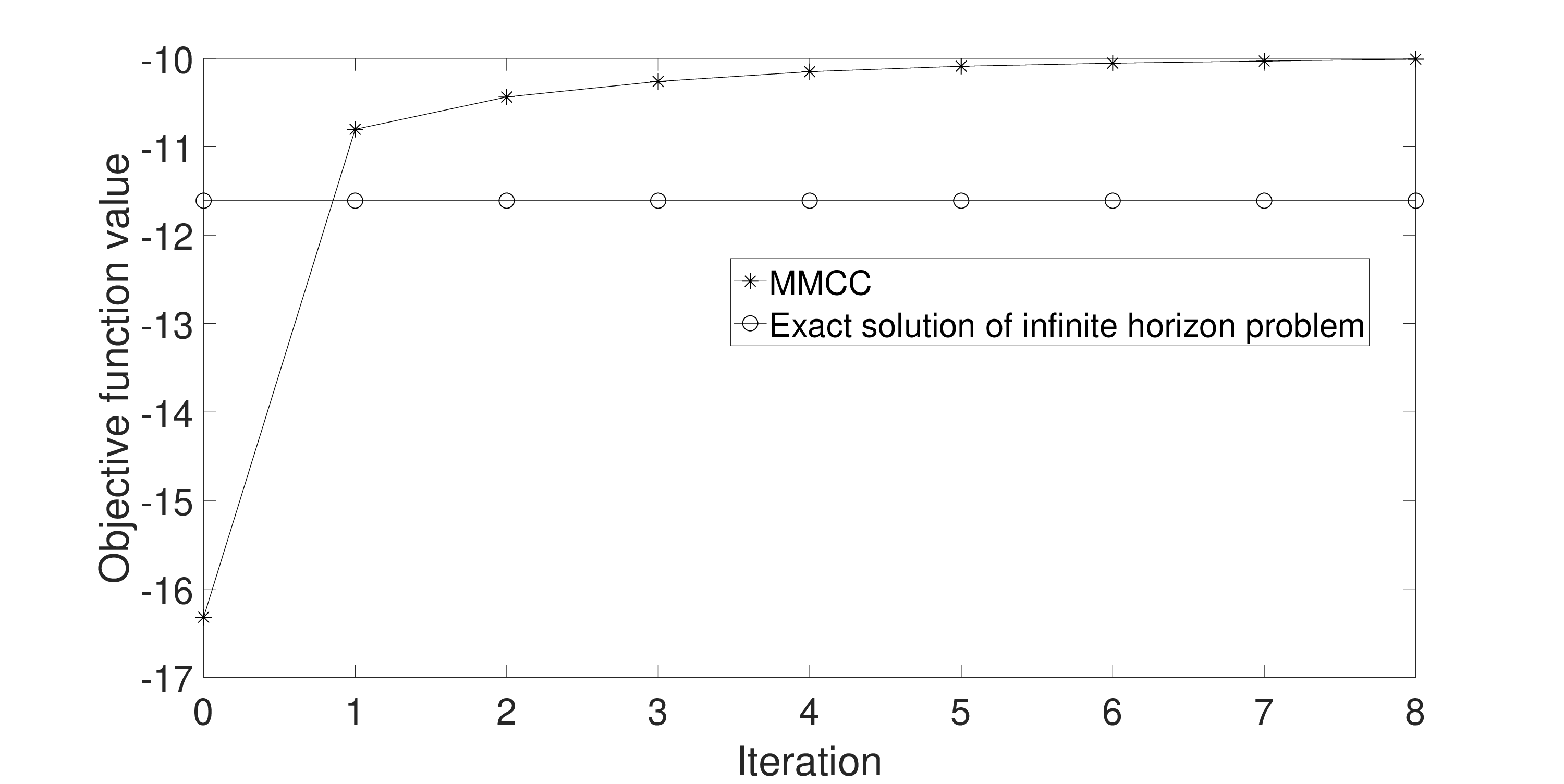}
	\end{centering}
	\caption{Objective function values of the MMCC algorithm defined in \eqref{equ:rbc_finite_obj} for the case of $T=5$. The MMCC algorithm converged after 9 iterations. It uses $N=19,200$ sample paths in the simulation and $m=300$ iterations in the Adam algorithm. The initial learning rate of the Adam algorithm is set to be 0.01. The minibatch size used in optimizing the neural network parameters is 64. It takes 4 minutes for each iteration under a Python implementation of the MMCC algorithm based on TensorFlow. The optimal objective value obtained by the MMCC algorithm is $-10.01$ (with a standard error of 0.024). The standard error is equal to the sample standard deviation of the $N$ samples of the objective function in \eqref{equ:rbc_finite_obj} divided by $\sqrt{N}$. The objective value obtained by the optimal solution for the infinite horizon problem is $-11.61$.\label{fig:RBC_opt_value_iter_T_5}}
\end{figure}

Figure \ref{fig:RBC_opt_value_iter_T_5} shows the objective function values of the MMCC algorithm for the case of $T=5$. The MMCC algorithm converged after 9 iterations. It uses $N=19,200$ sample paths in the simulation and $m=300$ iterations in the Adam algorithm. The initial learning rate of the Adam algorithm is set to be 0.01. The minibatch size used in optimizing the neural network parameters is 64. It takes 4 minutes for each iteration under a Python implementation of the MMCC algorithm based on TensorFlow. The optimal objective value obtained by the MMCC algorithm is $-10.01$ (with a standard error of 0.024). The standard error is equal to the sample standard deviation of the $N$ samples of the objective function in \eqref{equ:rbc_finite_obj} divided by $\sqrt{N}$. The objective function value obtained by the optimal solution for the infinite horizon problem is $-11.61$.

\begin{figure}[htb]
	\begin{centering}
		\includegraphics[width=\textwidth]{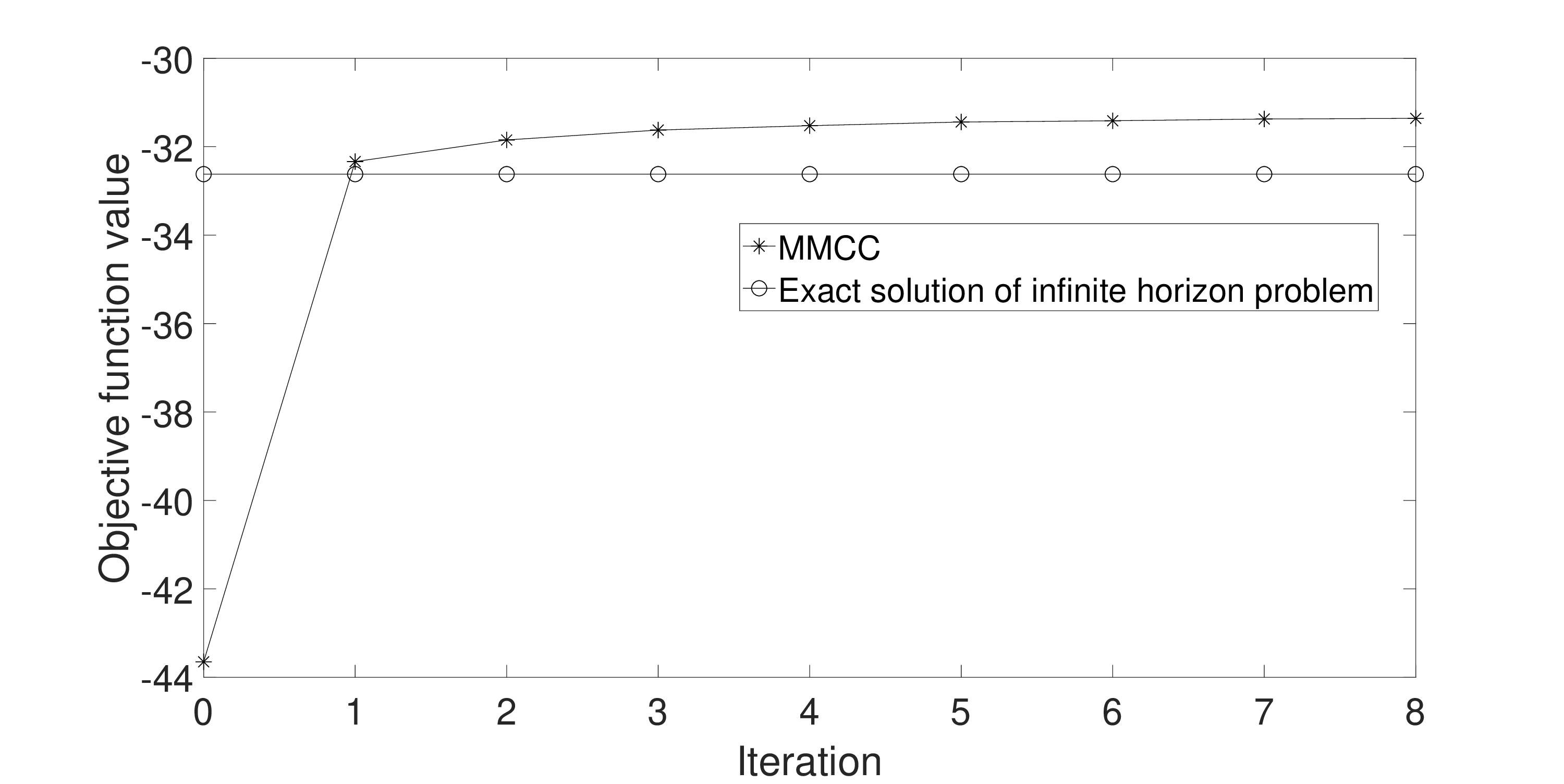}
	\end{centering}
	\caption{Objective function values of the MMCC algorithm defined in \eqref{equ:rbc_finite_obj} for the case of $T=10$. The MMCC algorithm converged after 9 iterations. It uses $N=19,200$ sample paths in the simulation and $m=300$ iterations in the Adam algorithm. The initial learning rate of the Adam algorithm is set to be 0.01. The minibatch size used in optimizing the neural network parameters is 64. It takes 18 minutes for each iteration under a Python implementation of the MMCC algorithm based on TensorFlow. The optimal objective value obtained by the MMCC algorithm is $-31.36$ (with a standard error of 0.045). The standard error is equal to the sample standard deviation of the $N$ samples of the objective function in \eqref{equ:rbc_finite_obj} divided by $\sqrt{N}$. The objective value obtained by the optimal solution for the infinite horizon problem is $-32.62$.\label{fig:RBC_opt_value_iter_T_10}}
\end{figure}

Figure \ref{fig:RBC_opt_value_iter_T_10} shows the objective function values of the MMCC algorithm for the case of $T=10$. 
The MMCC algorithm converged after 9 iterations. It takes 18 minutes for each iteration. The optimal objective value obtained by the MMCC algorithm is $-31.36$ (with a standard error of 0.045), while the objective value obtained by the optimal solution for the infinite horizon problem is $-32.62$.

\begin{figure}[htb]
	\begin{centering}
		\includegraphics[width=\textwidth]{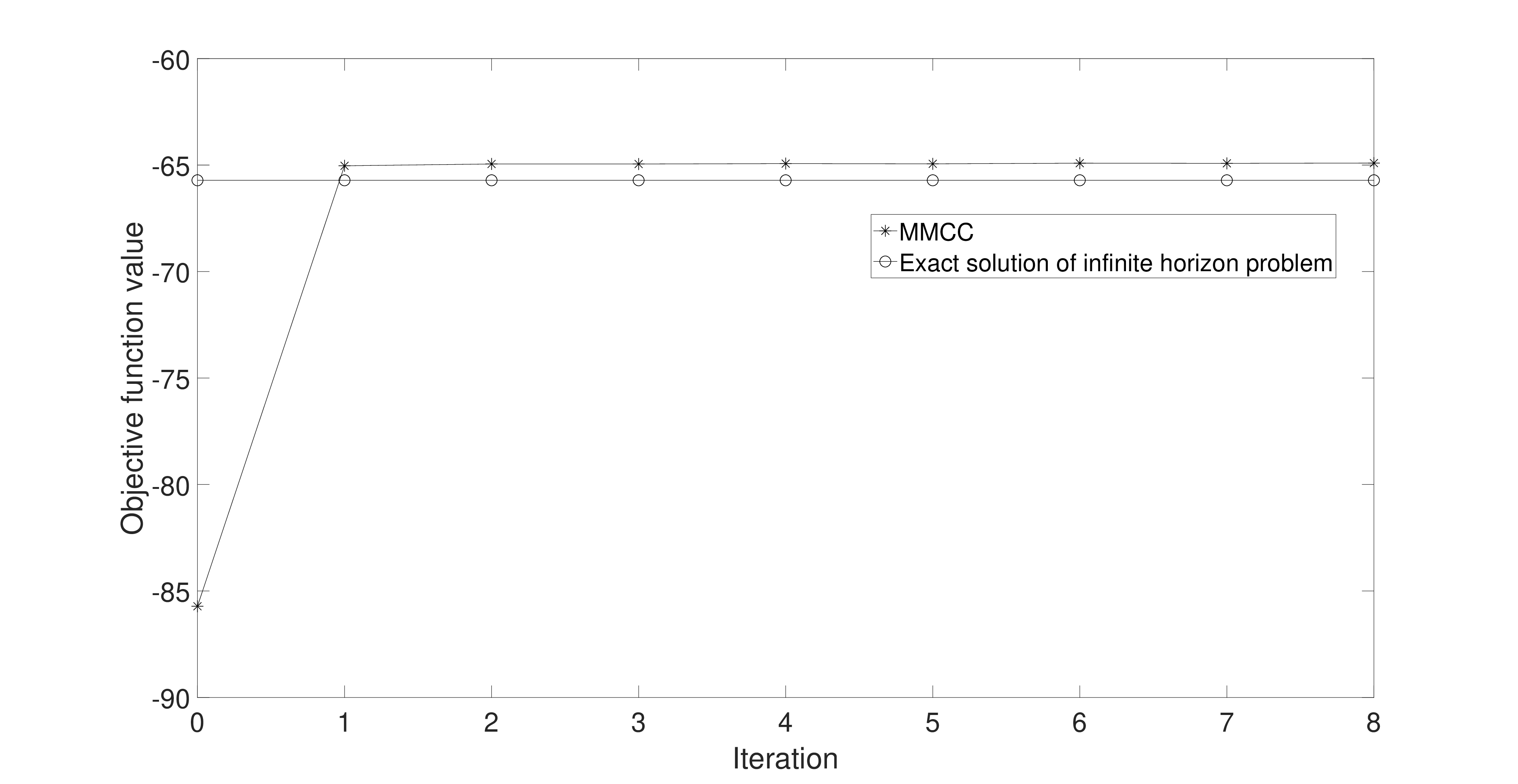}
	\end{centering}
	\caption{Objective function values of the MMCC algorithm defined in \eqref{equ:rbc_finite_obj} for the case of $T=20$. The MMCC algorithm converged after 3 iterations. It uses $N=19,200$ sample paths in the simulation and $m=300$ iterations in the Adam algorithm. The initial learning rate of the Adam algorithm is set to be 0.01. The minibatch size used in optimizing the neural network parameters is 64. It takes 78 minutes for each iteration under a Python implementation of the MMCC algorithm based on TensorFlow. The optimal objective value obtained by the MMCC algorithm is $-64.906$ (with a standard error of 0.029). The standard error is equal to the sample standard deviation of the $N$ samples of the objective function in \eqref{equ:rbc_finite_obj} divided by $\sqrt{N}$. The objective value obtained by the optimal solution for the infinite horizon problem is $-65.714$.\label{fig:RBC_opt_value_iter_T_20}}
\end{figure}

Figure \ref{fig:RBC_opt_value_iter_T_20} shows the objective function values of the MMCC algorithm for the case of $T=20$. 
The MMCC algorithm converged after 3 iterations. It takes 78 minutes for each iteration. The optimal objective value obtained by the MMCC algorithm is $-64.906$ (with a standard error of 0.029), while the objective value obtained by the optimal solution for the infinite horizon problem is $-65.714$.

\section{Application 3: Social Cost of Carbon}
\label{sec:scc}

\cite{Cai-Lontzek-2019} develop a framework of dynamic stochastic integration of climate and economy (DSICE) and show that the uncertainty about future economic and climate conditions substantially affects the choice of policies for managing the interaction between climate and economy. The DSICE model generalizes the commonly used dynamic integrated model of climate and the economy (DICE) model  \citep{Nordhaus-2008} by allowing for economic risks and climate risks.

\subsection{Problem Formulation}

The DSICE model consists of the climate model and the economic model. The climate model has three parts: the carbon system, the temperature system, and other climate conditions called tipping elements. 

There are two sources for carbon emissions at each time $t$: an industrial source, $E_{Ind,t}$, related to economic production activities, and 
an exogenous source, $E_{Land,t}$, arising from biological processes on the ground. The total emission at time $t$ is        	
The carbon concentration in the world at time $t$ is 
$M_{t} = (M_{AT,t},M_{UO,t},M_{LO,t})^{\top}$,
where the three components represent respectively the mass of carbon in the atmosphere, upper levels of the ocean, and lower levels of the ocean. 

The impact of carbon emissions on carbon concentration is represented by the linear dynamical system
\begin{align*}
M_{t+1}&=\Phi_{M}M_{t}+(E_{t},0,0)^{\top},\\
\Phi_{M}&=\begin{pmatrix}
1-\phi_{12} &  & \phi_{21} & 0\\
\phi_{12} & 1- & \phi_{21}-\phi_{23} & \phi_{32}\\
0 &  & \phi_{23} & 1-\phi_{32}
\end{pmatrix},
\end{align*}
where $\phi_{ij}$ is the rate at which carbon diffuses from level $i$ to level $j$, where $i, j=1, 2, 3$  
represent the atmosphere, upper ocean, and lower ocean, respectively. 

The temperature system at time $t$ consists of the temperatures of atmosphere $T_{AT}$ and the ocean $T_{OC}$, i.e., $T_{t}= (T_{AT,t}, T_{OC,t})^{\top}.$
The temperature system is governed by the diffusion of heat and evolves according to
\begin{align}
&T_{t+1}=\Phi_{T}T_{t}+(\xi_{1}\mathcal{F}_{t}(M_{AT,t}),0)^{\top},\\
&\Phi_{T}=\begin{pmatrix}1- \varphi_{21}-\xi_{2} & \varphi_{12} \\ \varphi_{21} & 1-\varphi_{12}
\end{pmatrix},
\end{align}
where $\phi_{ij}$ are the heat diffusion rates; $\xi_1$ represents coefficients of heating due to radioactive forcing; $\xi_2$ is the rate of cooling arising from infrared radiation to space. The
total radioactive forcing at time $t$ is
\begin{equation}
\mathcal{F}_{t}(M_{AT,t})=\eta\log_{2}(M_{AT,t}/M_{AT}^{*})+F_{EX,t},
\end{equation}
where $M_{AT}^{*}$ is the pre-industrial atmospheric carbon concentration, $\eta$ is the radioactive forcing parameter, and $F_{EX,t}$ is an exogenous process given by $F_{EX,t}=(-0.06+0.0036t)\cdot 1_{\{t\leq 100\}}+0.3\cdot 1_{\{t>100\}}.$

Tipping state $J$ represents some irreversible change in the climate system and is modeled by a discrete state Markov chain whose transition probabilities depend on the vector of climate states.  Specifically, $J_t$ stays at its initial value 0 until the tipping event is triggered, and then $J_t$ enters one of three discrete-state Markov chains $\{\mathcal{M}_1, \mathcal{M}_2, \mathcal{M}_3\}$ with equal probability.
	The transition process is represented as 
\begin{equation}
	J_{t+1}=g_{J}(T_{t},M_{t}, \omega_{J,t}),
\end{equation}
where $\omega_{J,t}$ is one serially independent stochastic process.

In the absence of climate damage, the gross world production is the Cobb-Douglas production function
\begin{align*}
	& f(K_t,L_t,\tilde{A}_{t})=\tilde{A}_{t}K_t^{\alpha}L_t^{1-\alpha},\\ 
	&\tilde{A}_{t}=\zeta_{t}A_{t}, A_{t}=A_{0}\exp(\alpha_{1}(1-e^{-\alpha_{2}t})/\alpha_{2}),
\end{align*}
where $K_{t}$ is the world capital stock at time $t$; 	$L_{t}=6514e^{-0.035t}+8600(1-e^{-0.035t})$ is the world population in millions at time $t$; 
$\alpha=0.3$ \citep*{Nordhaus-2008}; 
$\tilde{A}_{t}$ is productivity at time $t$, decomposed into the product of a deterministic trend $A_{t}$ and a stochastic productivity state $\zeta_{t}$. 
The DSICE model specifies a time-dependent, finite-state Markov chain for $(\zeta_{t},\chi_{t})$ with parameter values implying conditional and unconditional moments of consumption processes observed in market data. The Markov transition processes are denoted
\begin{align*}
	\zeta_{t+1}&=g_{\zeta}(\zeta_{t},\chi_{t}, \omega_{\zeta,t}), \ \ \ 	\chi_{t+1}=g_{\chi}(\chi_{t},\omega_{\chi,t}),
\end{align*}
where 
	$\omega_{\zeta,t}$, and $\omega_{\chi,t}$ are two serially independent stochastic processes.

In the DSICE model, the output 
is affected by the climate through the temperature $T_{AT}$ and the tipping state $J$. More precisely, the output under the impact of climate is assumed to be
\begin{align}
	& Y_{t}=\Omega(T_{AT,t},J_{t})f(K_{t},L_{t},\zeta_{t}A_{t}),\ \text{where}\\ 
	& \Omega(T_{AT,t},J_{t})=\Omega_{T}(T_{AT,t})\Omega_{J}(J_{t})=\frac{1}{1+\pi_{1}T_{AT,t}+\pi_{2}(T_{AT,t})^{2}}(1-D(J_{t})),	
\end{align}
in which $D(J_t)$ is the impact of tipping state $J$ on productivity.

The social planner can mitigate emissions by choosing a mitigation factor $\mu_{t}$, $0\leq\mu_{t} \leq 1$. Then, the industrial carbon emission at year $t$ equals
\begin{equation}
	E_{Ind,t}=\sigma_{t}(1-\mu_{t})f(K_{t},L_{t},\zeta_{t}A_{t}),\ \ \	\sigma_{t}=\sigma_{0}\exp\left(-0.0073\left(1-e^{-0.003t}\right)/0.003\right).
\end{equation}
Following \citet*{Nordhaus-2008}, the cost of mitigation level $\mu_{t}$ is
\begin{equation*}
	\Psi_{t}=\theta_{1,t}\mu_{t}^{\theta_{2}}Y_{t},\ \ \  	\theta_{1,t} = \frac{1.17\sigma_{t}(1+e^{-0.005t})}{2\theta_{2}}.
\end{equation*}
The world output net of damage is allocated across total consumption $C_{t}$, mitigation cost $\Psi_{t},$ and gross capital investment $I_{t}$:
\begin{equation}
	Y_{t}=C_{t}+\Psi_{t}+I_{t}.
\end{equation}
It is assumed that the capital stock evolves according to
\begin{equation}
	K_{t+1}=(1-\delta)K_{t}+I_{t},\ \ \ K_0=137 \text{ trillion dollars}.
\end{equation}
In the expected utility case of the DSICE model, the social planner seeks to solve the control problem
{\allowdisplaybreaks 
\begin{align}
V_{0}(S_0)=\max_{C_t,\mu_t,0\leq t\leq \mathcal{T}-1}\ \ & \mathbb{E}_0\left[\sum_{t=0}^{\mathcal{T}-1}\beta^t u(C_{t},L_{t})+\beta^\mathcal{T} V_\mathcal{T}(K_\mathcal{T}, M_\mathcal{T},  T_\mathcal{T})\right]\label{equ:DSICE_control_prob}\\
\text{s.t.}\quad\quad\ \ & Y_{t}=\Omega(T_{AT,t},J_{t})f(K_{t},L_{t},\zeta_{t}A_{t}), \notag\\
&\Psi_{t}=\theta_{1,t}\mu_{t}^{\theta_{2}}Y_{t},\notag\\
& K_{t+1}=(1-\delta)K_t+Y_{t}-C_{t}-\Psi_{t},\notag\\
& E_{t}= E_{Ind,t}+E_{Land,t},\notag\\
& E_{Ind,t}=\sigma_{t}(1-\mu_{t})f(K_{t},L_{t},\zeta_{t}A_{t}),\notag\\
&M_{t+1}=\Phi_{M}M_t+(E_{t},0,0)^{\top},\notag\\
&T_{t+1}=\Phi_{T}T_t+(\xi_{1}\mathcal{F}_{t}(M_{AT}),0)^{\top},\notag\\
&(\zeta_{t+1},\chi_{t+1})=(g_{\zeta}(\zeta_t,\chi_t,\omega_{\zeta, t}), g_{\chi}(\chi_t,\omega_{\chi,t})),\notag\\
&J_{t+1}=g_{J}(T_t,M_t,J_t,\omega_{J,t}),\notag
\end{align}}
where $\mathcal{T}=600$ years, $u(C_{t},L_{t})=\frac{(C_{t}/L_{t})^{1-1/\psi}}{1-1/\psi}L_{t}, t = 0, \ldots, \mathcal{T}-1$. The terminal value function is
\begin{equation}
	V_\mathcal{T}(K_{\mathcal{T}}, M_\mathcal{T},  T_\mathcal{T})=\sum_{t=\mathcal{T}}^{\mathcal{T}+\mathcal{S}}\beta^{t-\mathcal{T}} u(C_{t},L_{t}),
\end{equation}
where ${\mathcal{S}}=400$ years. It is assumed that after time $\mathcal{T}$, the system is deterministic,
population and productivity growth ends, all emissions are eliminated, and the consumption-output ratio is fixed at $0.78$.

The control policy of the problem \eqref{equ:DSICE_control_prob} needs to satisfy the constraints
\begin{align}\label{equ:DSICE_policy_constr}
	0&\leq \mu_t \leq 1,\\
	0&\leq I_t=Y_t-C_t-\Psi_t=(1-\theta_{1,t}\mu_t^{\theta_2})Y_t-C_t,
\end{align}
which implies that 
\begin{equation}\label{equ:DSICE_C_constr}
	0\leq C_t\leq (1-\theta_{1,t}\mu_t^{\theta_2})Y_t.	
\end{equation}
To impose the constraint in \eqref{equ:DSICE_C_constr}, we introduce a new control variable $p_t$ and represent $C_t$ as 
\begin{align}
	C_t&=p_t (1-\theta_{1,t}\mu_t^{\theta_2})Y_t,\ \text{where}\\
	0&\leq p_t\leq 1.
\end{align}
Then, we will parameterize the control policy $c_t=(\mu_t, p_t)$ by neural networks in the MMCC algorithm.

The control problem \eqref{equ:DSICE_control_prob} is difficult to solve as it is a high-dimensional problem with a large number of time periods and a nine-dimensional state vector $S_t=(K_t,M_t,T_t,\zeta_t,\chi_t,J_t)$.
In this formulation, the tipping state variable $J_{t+1}$ is specified as a discrete random variable whose distribution depends on the control variable $\mu_t$ but $J_{t+1}$ is not differentiable with respect to 
$\mu_t$. 
Consequently, we implement the MMCC algorithm under the assumption of no tipping events and leave the case with tipping events for future research.

\subsection{Numerical Results}
 In this numerical example, we specify the parameters (such as $\Phi_M$, $\Phi_T$, $M_{AT}^*$, $\beta$, $\alpha_1$, $\alpha_2$, $\theta_1$, and $\theta_2$), dynamics (such as $\omega_{J,t}$, $\omega_{\zeta,t}$, and $\omega_{\chi,t}$), and functions (such as $g_J$, $g_\zeta$, $g_\chi$, and $f$) to be the same as those in \citet*{Cai-Lontzek-2019}.

We specify the neural network for $c_t=(\mu_t, p_t)$ 
to have seven layers, where the input layer and the output layer respectively have 8 and 2 neurons, and the five hidden layers have 100 neurons.  
The nonlinear activation functions of the hidden layer and the output layer are respectively the rectified linear function and two sigmoid functions.
In total, there are $T-1$ neural networks corresponding to $c_t$, $t=1, 2, \ldots, T-1$.

To speed up the calculation, we split the $T-1$ neural networks into 6 groups, each containing 100 neural networks except one containing 99 neural networks. The parameters in each group are optimized at the same time. The minibatch size in Adam is 128.
We use $N=25,600$ sample paths in the simulation and $m=200$ iterations in the Adam algorithm.

\begin{figure}[!htbp]
\begin{centering}
	\includegraphics[width=\textwidth]{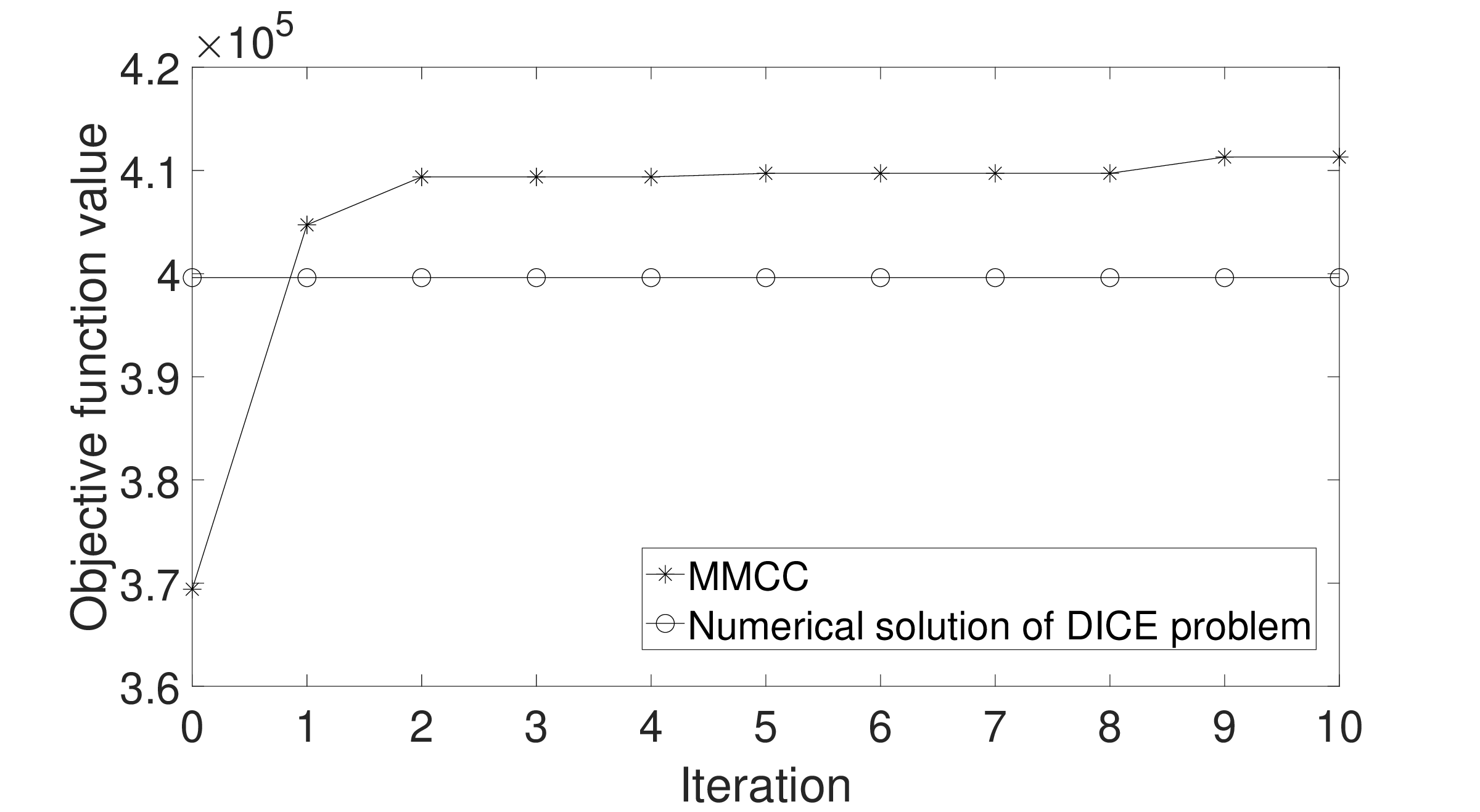}
\end{centering}
    \caption{Objective function values of the MMCC algorithm. The algorithm is implemented by TensorFlow in Python. It takes less than 43 hours to converge on a computer node with 2 Intel Xeon E5-2697A V4 CPUs (2.6G hz), each having 16 cores. The optimal objective value obtained by the MMCC algorithm is $411,302.6$ (with a standard error of $1396.3$). The numerical solution of the DICE problem (deterministic version of the problem) is $399,614.0$.}
    \label{fig:SCC_iterations}
\end{figure}

Figure \ref{fig:SCC_iterations} shows the iteration of the MMCC algorithm for solving the problem \eqref{equ:DSICE_control_prob}. The algorithm is implemented by TensorFlow in Python.  It takes less than 43 hours to converge on a computer node with 2 Intel Xeon E5-2697A V4 CPUs (2.6G hz), each having 16 cores. The optimal objective value obtained by the MMCC algorithm is $411,302.6$ (with a standard error of $1396.3$). For a rough comparison, the numerical solution of the DICE problem is $399,614.0$. The numerical solution is computed in DICE-CJL \citep*{cai2012continuous}, which changes the original DICE model with  
10-year time periods to be a model with the one-year time period, which is consistent with the problem formulation in \eqref{equ:DSICE_control_prob}.

\begin{appendices}

	\section{A Simple Derivation}\label{app:simple_deriv}
	We will show that \eqref{eq:monot} is equivalent to \eqref{equ:new_2}. In fact, by \eqref{equ:utility_func}, \eqref{eq:monot} is equivalent to
	{\allowdisplaybreaks
		\begin{align}\label{equ:new_1}
		& \mathbb{E}_0\left[\sum_{j=0}^{t-1} u_{j+1}(s_{j+1}, s_{j}, c_j)\right.\notag\\
		&\quad\quad\quad\quad\quad \left.+\sum_{j=t}^{T-1} u_{j+1}(s_{j+1},s_{j},c_j)\middle | c_0^{k-1}, \theta_1^{k-1}, \ldots, \theta_{t-1}^{k-1}, \theta_t^k, \theta_{t+1}^k, \ldots, \theta_{T-1}^k \right]\notag\\
		\geq{} & \mathbb{E}_0\left[\sum_{j=0}^{t-1} u_{j+1}(s_{j+1}, s_{j}, c_j)\right.\notag\\
		&\quad\quad\quad\quad\quad \left.+\sum_{j=t}^{T-1} u_{j+1}(s_{j+1},s_{j},c_j)\middle | c_0^{k-1}, \theta_1^{k-1}, \ldots, \theta_{t-1}^{k-1}, \theta_t^{k-1}, \theta_{t+1}^k, \ldots, \theta_{T-1}^k \right].
		\end{align}}
	By \eqref{eq:c_t_s_t} and \eqref{eq:state_evo}, $\sum_{j=0}^{t-1} u_{j+1}(s_{j+1}, s_{j}, c_j)$ depends on the control parameters $(c_0, \theta_1, \ldots, \theta_{t-1})$ but not on the control parameter $(\theta_{t}, \ldots, \theta_{T-1})$. Therefore, we have
	{\allowdisplaybreaks
		\begin{align*}
		&  \mathbb{E}_0\left[\sum_{j=0}^{t-1} u_{j+1}(s_{j+1}, s_{j}, c_j)\middle | c_0^{k-1}, \theta_1^{k-1}, \ldots, \theta_{t-1}^{k-1}, \theta_t^k, \theta_{t+1}^k, \ldots, \theta_{T-1}^k \right]\\
		={} & \mathbb{E}_0\left[\sum_{j=0}^{t-1} u_{j+1}(s_{j+1}, s_{j}, c_j)\middle | c_0^{k-1}, \theta_1^{k-1}, \ldots, \theta_{t-1}^{k-1}, \theta_t^{k-1}, \theta_{t+1}^k, \ldots, \theta_{T-1}^k \right],
		\end{align*}}
	which implies that \eqref{equ:new_1} is equivalent to \eqref{equ:new_2}.
	
	\section{Proof of Theorems}
	\subsection{Proof of Theorem \ref{thm:monoto}}\label{app:proof_monoto}
	\begin{proof}
		In the MMCC algorithm, the iterations satisfy 
		\eqref{eq:monot} and \eqref{eq:mono_0}. Therefore, we have
		{\allowdisplaybreaks
			\begin{align}
			& \phantom{{}={}} U(c^{k-1}_0, \theta^{k-1}_1, \theta^{k-1}_2, \ldots, \theta^{k-1}_{T-3}, \theta^{k-1}_{T-2}, \theta^{k-1}_{T-1})\notag\\
			&\leq  U(c^{k-1}_0, \theta^{k-1}_1, \theta^{k-1}_2, \ldots, \theta^{k-1}_{T-3}, \theta^{k-1}_{T-2}, \theta^{k}_{T-1})\notag\\
			&\leq  U(c^{k-1}_0, \theta^{k-1}_1, \theta^{k-1}_2, \ldots, \theta^{k-1}_{T-3}, \theta^{k}_{T-2}, \theta^{k}_{T-1})\notag\\
			&\leq  \cdots \notag\\
			&\leq  U(c^{k-1}_0, \theta^{k}_1, \theta^{k}_2, \ldots, \theta^{k}_{T-3}, \theta^{k}_{T-2}, \theta^{k}_{T-1})\notag\\
			&\leq  U(c^{k}_0, \theta^{k}_1, \theta^{k}_2, \ldots, \theta^{k}_{T-3}, \theta^{k}_{T-2}, \theta^{k}_{T-1}),\notag
			\end{align}}
		from which the proof is completed.
	\end{proof}
	
	\subsection{Proof of Theorem \ref{thm:convergence_red_em}}\label{app:proof_red_em}
	\begin{proof}
		We first recall the following definition in \citet*{wu1983convergence}: A point-to-set map $\rho$ on $X$ is said to be closed at $x$, if $x^k \to x$, $x^k\in X$, $y^k \to y$, and $y^k\in \rho(x^k)$ imply $y\in \rho(x)$.
        
		We also recall the following global convergence theorem (\citet[][p. 91]{Zangwill1969}): Let the sequence $\{x^k\}_{k=0}^{\infty}$ be generated by $x^k\in M(x^{k-1})$, where $M$ is a point-to-set map on $X$. Let a solution set $\Gamma \subset X$ to be given, and suppose that: (i) all points $x^k$ are contained in a compact set $S \subset X$; (ii) $M$ is closed over the complement of $\Gamma$; (iii) there is a continuous function $\alpha$ on $X$ such that (a) if $x \notin \Gamma$, $\alpha(y) > \alpha(x)$ for all $y \in M(x)$, and (b) if $x \in \Gamma$, $\alpha(y) \geq \alpha(x)$ for all $y \in M(x)$. Then all the limit points of ${x^k}$ are in the solution set $\Gamma$ and $\alpha(x^k)$ converges monotonically to $\alpha(x^*)$ for some $x^*\in \Gamma$.
		
		We now prove part (1) of the theorem. First, we show that $M$ is a closed point-to-set map on $\R^n$. Suppose
		$$a^k=(a_0^k, a_1^k, \ldots, a^k_{T-1})\to \bar a=(\bar a_0, \bar a_1, \ldots, \bar a_{T-1}),\ \text{as}\ k\to\infty.$$
		Suppose $b^k=(b_0^k, b_1^k, \ldots, b^k_{T-1})\in M(a^k)$ and $b^k\to \bar b=(\bar b_0, \bar b_1, \ldots, \bar b_{T-1})$ as $k\to\infty$. We will show that $\bar b\in M(\bar a)$. Since $b^k\in M(a^k)$, it follows that
		{\allowdisplaybreaks
			\begin{align}
			&U(a_0^k, a_1^k, \ldots, a^k_{T-2}, b^k_{T-1})\geq U(a_0^k, a_1^k, \ldots, a^k_{T-2}, a^k_{T-1}), \forall k\notag\\
			&U(a_0^k, a_1^k, \ldots, a^k_{t-1}, b^k_{t}, b^k_{t+1}, \ldots, b^k_{T-1})\geq U(a_0^k, a_1^k, \ldots, a^k_{t-1}, a^k_{t}, b^k_{t+1}, \ldots, b^k_{T-1}), \forall t, \forall k\notag\\
			&U(b_0^k, b_1^k, \ldots, b^k_{T-1})\geq U(a_0^k, b_1^k, \ldots, b^k_{T-1}), \forall k.\notag
			\end{align}}
		Letting $k\to \infty$ in the above inequalities, we obtain from the continuity of $U$ that
		{\allowdisplaybreaks
			\begin{align}
			&U(\bar a_0, \bar a_1, \ldots, \bar a_{T-2}, \bar b_{T-1})\geq U(\bar a_0, \bar a_1, \ldots, \bar a_{T-2}, \bar a_{T-1}), \forall k\notag\\
			&U(\bar a_0, \bar a_1, \ldots, \bar a_{t-1}, \bar b_{t}, \bar b_{t+1}, \ldots, \bar b_{T-1})\geq U(\bar a_0, \bar a_1, \ldots, \bar a_{t-1}, \bar a_{t}, \bar b_{t+1}, \ldots, \bar b_{T-1}), \forall t, \forall k\notag\\
			&U(\bar b_0, \bar b_1, \ldots, \bar b_{T-1})\geq U(\bar a_0, \bar b_1, \ldots, \bar b_{T-1}), \forall k,\notag
			\end{align}}
		which implies that $\bar b\in M(\bar a)$. Hence, $M$ is a closed point-to-set map on $\R^n$.
		
		Second, we will verify that the conditions of the global convergence theorem cited above hold. Let $\alpha(x)$ be $U(x)$ and the solution set $\Gamma$ to be $\mathcal{S}$ or $\mathcal{M}$. Then, condition (i) follows from \eqref{equ:assump_1} and \eqref{equ:monoto}. Condition (ii) has been approved above. Condition (iii) (a) follows from \eqref{equ:cond_converg}. Condition (iii) (b) follows from \eqref{equ:monoto}. Hence, the conclusion of part (1) of the theorem follows from the global convergence theorem.
		
		We move to prove part (2) of the theorem. To prove part (2), we only need to show that, under the condition of part (2), \eqref{equ:cond_converg} holds for any $x^{k-1}\notin \mathcal{S}$. For any such $x^{k-1}$, it follows from
		the definition of the set $\mathcal{S}$ that
		$\frac{\partial U(x^{k-1})}{\partial x^{k-1}}\neq 0$. Suppose $x^k = x^{k-1}$. Then, for each $j=T-1, T-2, \ldots, 1, 0$, $x_j^{k-1}$ maximizes the function $H_j(y):=U(x_0^{k-1}, x_1^{k-1}, \ldots, x_{j-1}^{k-1}, y, x_{j+1}^{k-1}, \ldots, x_{T-1}^{k-1})$, which implies that $\frac{\partial U(x^{k-1})}{\partial x_j^{k-1}}=0$ for all $j$, which contradicts to that $\frac{\partial U(x^{k-1})}{\partial x^{k-1}}\neq 0$. Hence, $x^k \neq x^{k-1}$. Let $i_0$ be the largest index $j\in\{0, 1, \ldots, T-1\}$ such that $x^k_j\neq x^{k-1}_j$. Then, by the specification of the algorithm,
		$x_{i_0}^{k}$ maximizes the function $H_{i_0}(y):=U(x_0^{k-1}, x_1^{k-1}, \ldots, x_{i_0-1}^{k-1}, y, x_{i_0+1}^{k-1}, \ldots, x_{T-1}^{k-1})$ but $x_{i_0}^{k-1}$ does not. Hence,
		$$H_{i_0}(x_{i_0}^k)>H_{i_0}(x_{i_0}^{k-1})=U(x^{k-1}),$$
		which implies that
		$$U(x^{k})\geq H_{i_0}(x_{i_0}^k)>U(x^{k-1}).$$
		Hence, \eqref{equ:cond_converg} holds for any $x^{k-1}\notin \mathcal{S}$ for the MMCC algorithm. Then, the conclusion of part (2) follows from part (1) of the theorem, which has been proved.
	\end{proof}
	
	\subsection{Proof of Theorem \ref{thm:converg_x}}
	\label{app:proof_cong_x}
	\begin{proof}
		We first prove part (1). By Theorem \ref{thm:convergence_red_em}, all the limit points of $\{x^k\}_{k\geq 0}$ are in $\cS(U^*)=\{x^*\}$ (resp. $\M(U^*)=\{x^*\}$). Hence, any converging subsequence of $\{x^k\}_{k\geq 0}$ converges to $x^*$, which implies that $x^k\to x^*$ as $k\to\infty$. Hence, part (1) of the theorem holds.
		Next, we prove part (2). By the condition \eqref{equ:assump_1}, $\{x^k\}$ is a bounded sequence. By Theorem 28.1 of \citet*{Ostrowski-1966}, the set of limit points of the bounded sequence $\{x^k\}$ with $\|x^{k+1}-x^k\|\to 0$ as $k\to\infty$ is compact and connected. In addition, by Theorem \ref{thm:convergence_red_em}, all the limit points of $\{x^k\}$ are in $\cS(U^*)$ (resp. $\M(U^*)$). Hence, the conclusion of part (2) follows.
	\end{proof}

    \section{The MMCC Algorithm for the General Control Problem}
	\label{app:CEM-general-control}
	
	The MMCC algorithm also works for the general control problem \eqref{equ:multi_per_obj_gen} in which the utility function may not be time-separable. The convergence theorems \ref{thm:monoto}, \ref{thm:convergence_red_em}, and \ref{thm:converg_x} also hold for the MMCC algorithm for the general control problem \eqref{equ:multi_per_obj_gen}. However, in such cases, \eqref{eq:monot} can no longer be simplified to be \eqref{equ:new_2}. 
    
    We generalize the Algorithm \ref{alg:Main_Algorithm} to be the Algorithm \ref{alg:General-Algorithm} for solving problem \eqref{equ:multi_per_obj_gen}.

\begin{algorithm}[htpb]
\caption{The MMCC algorithm for solving problem \eqref{equ:multi_per_obj_gen}.}
\label{alg:General-Algorithm}
\begin{enumerate}
\item Initialize $k=1$. Choose the initial policy $x^0:=(c^0_0, \theta^0_{1}, \theta^0_{2}, \ldots, \theta^0_{T-1})$. 
\item Iterate $k$ until some stopping criteria are met. In the $k$th iteration, update $x^{k-1}=(c_0^{k-1}, \theta_1^{k-1}, \theta_2^{k-1}, \ldots, \theta_{T-1}^{k-1})$ to $x^k=(c_0^{k}, \theta_1^{k}, \theta_2^{k}, \ldots, \theta_{T-1}^{k})$ by moving backwards from $t=T-1$ to $t=0$ as follows:
\end{enumerate}
 \begin{enumerate}

   \item[(a)] Move backward from $t=T-1$ to $t=1$. At time $t$, update $\theta_{t}^{k-1}$ by $\theta_{t}^{k}$ such that
           \begin{align}
           &\mathbb{E}_0\left[ \left. u\left( s_0,c_0,s_1,c_1,\ldots,s_{T-1},c_T,s_T \right) \right|c_{0}^{k-1},\theta _{1}^{k-1},\ldots,\theta _{t-1}^{k-1},\theta _{t}^{k},\theta _{t+1}^{k},\ldots,\theta _{T-1}^{k} \right] \notag
            \\
            &\geq \mathbb{E}_0\left[ \left. u\left( s_0,c_0,s_1,c_1,\ldots,s_{T-1},c_T,s_T \right) \right|c_{0}^{k-1},\theta _{1}^{k-1},\ldots,\theta _{t-1}^{k-1},\theta _{t}^{k-1},\theta _{t+1}^{k},\ldots,\theta _{T-1}^{k} \right].\label{eq:gen_opt_time_t_revise_general_sim}
           \end{align}
           Such $\theta^{k}_{t}$ can be set as a suboptimal (optimal) solution to the problem
            \begin{equation}
	\label{eq:opt_t_S_gen}
	\max_{\theta_{t}\in \Theta_t}
	\mathbb{E}_0\left[u(s_0, c_0, s_1, c_1, \ldots, s_{T-1}, c_{T-1}, s_{T})\middle | c_0^{k-1}, \theta_1^{k-1}, \ldots, \theta_{t-1}^{k-1}, \theta_t, \theta_{t+1}^k, \ldots, \theta_{T-1}^k\right],
	\end{equation}
 where $\Theta_t=\{\theta\in \R^d\mid (c_0^{k-1}, \theta_1^{k-1}, \ldots, \theta_{t-1}^{k-1}, \theta, \theta_{t+1}^k, \ldots, \theta_{T-1}^k)\in\Theta\}$.
		
    \item[(b)] At period $0$, update $c_{0}^{k-1}$ to be $c_{0}^{k}$ such that
    \begin{align}
       &\mathbb{E}_0\left[ \left. u\left( s_0,c_0,s_1,c_1,\ldots,s_{T-1},c_T,s_T \right) \right|c_{0}^{k},\theta _{1}^{k},\ldots,\theta _{T-1}^{k} \right] \notag
\\
&\geq \mathbb{E}_0\left[ \left. u\left( s_0,c_0,s_1,c_1,\ldots,s_{T-1},c_T,s_T \right) \right|c_{0}^{k-1},\theta _{1}^{k},\ldots,\theta _{T-1}^{k} \right] .\label{eq:gen_opt_time_0_general}
       \end{align}
    Such $c^{k}_0$ can be set as a suboptimal (optimal) solution to the problem
    \begin{align}
   \max_{c_0\in \Theta _0} \mathbb{E}_0\left[ u\left( s_0,c_0,s_1,c_1,\ldots,s_{T-1},c_T,s_T \right) \middle| c_0,\theta _{1}^{k},\ldots,\theta _{T-1}^{k} \right],\label{eq:opt_t_S_gen_t_0}
    \end{align}
    where $\Theta_0=\{c\in\mathbb{R}^{n_c}\mid (c,\theta_1^{k}, \ldots, \theta_{T-1}^{k})\in\Theta \}$.
    \end{enumerate}
\end{algorithm}
\end{appendices}

\vspace{1em}

{\bf ACKNOWLEDGMENT.} We are grateful to Paul Glasserman for his insightful comments and helpful discussions.
We also thank seminar and conference participants at Harvard University, Hong Kong Polytechnic University, INFORMS 2016 Annual Meeting, SIAM Financial Mathematics Conference 2016,  QMF Conference 2016, and The workshop on Machine Learning and FinTech at NUS in 2019 for their useful comments. Xianhua Peng is partially supported by the Natural Science Foundation of Shenzhen (Grant No. JCYJ20190813104607549) and the National Natural Science Foundation of China (Grant No. 72150003).

\newpage
\bibliographystyle{dcu}
\bibliography{dynamic_EM_ref, new_ref}

\end{document}